\documentclass[review]{elsarticle}

\makeatletter
\def\bs{\expandafter\@gobble\string\\}
\def\lb{\expandafter\@gobble\string\{}
\def\rb{\expandafter\@gobble\string\}}
\def\@pdfauthor{M.G.C. }
\def\@pdftitle{elsarticle.cls -- A documentation}
\def\@pdfsubject{Document formatting with elsarticle.cls}
\def\@pdfkeywords{LaTeX, Elsevier Ltd, document class}

 \usepackage{graphics}
\usepackage{epsfig}
\usepackage{hyperref}
\usepackage{tabularx}

\usepackage{amssymb}
\usepackage{amsthm}
\usepackage{algorithm}
\usepackage{multirow}
\usepackage[noend]{algpseudocode}
\usepackage{bm}
\usepackage{color}
\usepackage{import}
\usepackage{tikz}
\usepackage{subcaption}

\makeatletter
\@addtoreset{subfigure}{row}
\makeatother
\usepackage{siunitx}
\usetikzlibrary{arrows, decorations.markings}
      \tikzstyle{vecArrow} = [thick, decoration={markings,mark=at position
      1 with {\arrow[semithick]{open triangle 60}}},
      double distance=1.4pt, shorten >= 5.5pt,
      preaction = {decorate},
      postaction = {draw,line width=1.4pt, white,shorten >= 4.5pt}]
     \tikzstyle{innerWhite} = [semithick, white,line width=1.4pt, shorten >= 4.5pt]

\usetikzlibrary{positioning,shapes.geometric}

\newcommand{\vertSpacing}{10mm}
\newcommand{\textHeight}{1.cm}

\tikzset{
	block step/.style={
		rectangle, draw, text width=5cm, inner sep=0.1cm,
		minimum height=\textHeight, text centered, font=\sffamily,
		rounded corners
	},
	block step2/.style={
		rectangle, draw, text width=4cm, inner sep=0.1cm,
		minimum height=\textHeight, text centered, font=\sffamily,
		rounded corners
	},
	block picardstep/.style={
		rectangle, draw, text width=5cm, inner sep=0.1cm,
		minimum height=\textHeight, text centered, font=\sffamily
	},
	block picardstep2/.style={
		rectangle, draw, text width=8cm, inner sep=0.1cm,
		minimum height=\textHeight, text centered, font=\sffamily
	},
	block decision/.style={
		block picardstep=#1, text width=4cm, diamond, aspect=2
	},
	block picard/.style={
		block picardstep=#1, circle, inner sep=0pt, text height=0.1cm,
		text width=0.1cm, minimum height=0.2cm
	}
}

\newcounter{bla}

\newcommand{\beq}{\begin{equation}}
\newcommand{\eeq}{\end{equation}}
\newcommand{\beqs}{\begin{equation*}}
\newcommand{\eeqs}{\end{equation*}}
\usepackage{amsmath}
\newtheorem*{remark}{Remark}

\begin{document}

\begin{frontmatter}

\title{An immersed boundary method for fluid-structure interaction based on overlapping domain decomposition}

\author[a]{Maria Giuseppina Chiara Nestola \corref{author}\sep\fnref{author2}}
\author[b]{Barna Becsek\fnref{author2}}
\author[b]{Hadi Zolfaghari}
\author[a]{Patrick Zulian}
\author[b]{Dario De Marinis}
\author[a]{Rolf Krause}
\author[b]{Dominik Obrist}

\cortext[author] {Corresponding author. \textit{E-mail address:} nestom@usi.ch}
\fntext[author2] {Both authors contributed equally to the manuscript}
\address[a]{Institute of Computational Science, Center for Computational Medicine in Cardiology (CCMC), Universit\`a della Svizzera Italiana, Via Giuseppe Buffi 13, 9600 
Lugano, Switzerland}
\address[b]{ARTORG Center for Biomedical Engineering Research, University of Bern,
Murtenstrasse 50, 3008 Bern, Switzerland}

\begin{abstract}
We present a novel framework inspired by the Immersed Boundary Method for predicting the fluid-structure interaction of complex structures immersed in flows with moderate to high Reynolds numbers. 

The main novelties of the proposed fluid-structure interaction framework are 1) the use of elastodynamics equations for the structure, 
2) the use of a high-order Navier--Stokes solver for the flow, and 3) the variational transfer ($L^2$-projection) for coupling the solid and fluid subproblems.

The dynamic behavior of a deformable structure is simulated in a finite element framework by adopting a fully implicit scheme for its 
temporal integration. It allows for mechanical constitutive laws including nonhomogeneous and fiber-reinforced materials. 

The Navier--Stokes equations for the incompressible flow are discretized with high-order finite differences which allow for the direct 
numerical simulation of laminar, transitional and turbulent flows. 

The structure and the flow solvers are coupled by using an $L^2$-projection method for the transfer of velocities and forces between the fluid grid and the solid mesh. This strategy allows for the numerical solution of coupled large scale problems based on nonconforming structured and unstructured grids. 

The framework is validated with the Turek--Hron benchmark and a newly proposed benchmark
modelling the flow-induced oscillation of an inert plate. A three-dimensional simulation of an elastic beam in transitional flow is 
provided to show the solver's capability of coping with anisotropic elastic structures immersed in complex fluid flow.
\end{abstract}


\end{frontmatter}

\noindent

\section{Introduction}                                          
\label{Intro}
Over the past decades, Fluid-Structure Interaction (FSI)  \cite{peskin1972flow, peskin1972, liu2006}  analysis of the cardiovascular
system and, in particular, of heart valves has become an increasingly active area of research 
\cite{kamensky2014variational, de2016moving, griffith2009simulating, mcgee2016computational,
de2009direct, nobili2008numerical}. 
The main difficulties related to numerical simulation of FSI problems are: (a) the existence of a two-field problem where the two phases (i.e. fluid and structure) are
separated by a common boundary whose position is an unknown of the problem (geometrical nonlinearity); 
(b) the treatment of  the interface conditions ensuring the continuity of the velocity and the stress across the 
interface \cite{donea2003finite}; (c) the interaction with thin and/or bulky solid structures which may exhibit large 
deformations, and (d) the simulation of moderate- to high-Reynolds-number flows involving transition
from laminar to turbulent flow. 
Moreover, high-fidelity simulations of complex and large-scale problems, such as the interaction 
between blood flow and heart valves, demand for the development of high-performance numerical libraries. Such libraries are  
optimized for modern supercomputers by ensuring a high level of parallelism, scalability, flexibility, and efficiency.

In literature, several approaches have been developed for FSI simulations, which can be classified in boundary-fitted 
\cite{donea2004arbitrary} and embedded-boundary methods 
\cite{mittal2005immersed, Zhang2007}. 

In boundary-fitted methods, the fluid subproblem is solved in a moving spatial domain where the Navier--Stokes 
equations are formulated in an Arbitrary Lagrangian Eulerian framework \cite{donea2004arbitrary, Nestola2016} while the solid structure is usually analyzed in a 
Lagrangian fashion. Although this approach is known to produce accurate results at the 
interface between solid and fluid, the fluid grid may become severely distorted for scenarios that involve large displacements and/or rotations,
such that the numerical stability of the coupled problem and the accuracy of the solution can be affected. In particular, in heart valve simulations, it can become 
computationally very expensive to preserve good mesh quality because the movement of the 
valve leaflets and their contact during valve closure change the topology of the fluid domain. 

In order to circumvent those difficulties, embedded-boundary approaches such as the Immersed Boundary Method (IBM) have 
been introduced for simulating the complex dynamics of the heart. The main characteristic of this 
approach is the representation of the immersed structure by a force density term in the 
Navier--Stokes equations. 

In the original IBM Peskin \cite{peskin1972} adopted a finite difference scheme for the spatial 
discretization of the fluid subproblem and a Lagrangian model with one-dimensional fiber-like 
elements for the structure.  The solid and the fluid subproblems were coupled by 
interaction equations involving a smoothed approximation of the Dirac-delta function to interpolate 
data between Eulerian flow and Lagrangian structure variables. 

Since the original development of this method by Peskin, a large number of modified approaches were proposed 
to simulate flow over geometries on nonconforming grids.

Devendran and Peskin \cite{devendran2012} developed an energy functional based version 
of the conventional IBM that allows for a nodal approximation of the elastic forces generated by an immersed 
hyperelastic material via a finite element type approximation. Griffith et al. \cite{griffith2012hybrid} introduced a version of the IBM describing the solid body 
motion via standard Lagrangian finite element methods. Rather than spreading forces from the nodes of the Lagrangian mesh and interpolating velocities to those mesh nodes, forces are spread from (and velocities are interpolated to) dynamically selected quadrature points 
defined within the Lagrangian structural elements.

Other approaches include the potential embedded method whose main idea is modelling the structure via a potential 
energy and the sharp-interface methodology \cite{mittal2008versatile, sotiropoulos2009review} where the use of a 
multidimensional ghost-cell technique allows to satisfy the boundary conditions precisely, avoiding spurious 
spreading of boundary forcing into the fluid.

In the Immersed Finite Element Method (IFEM) the discretization of both the fluid and the solid 
subproblems are formulated in a finite element fashion \cite{Boffi2003, Glowinski2007, Zhang2007}. 
Glowinski et al. \cite{Glowinski2007} adopted the reproducing kernel particle 
method (RKPM) to approximate the Dirac-delta distribution for interpolating the fluid velocities from 
Eulerian (fluid) to Lagrangian (structure) coordinates 
and spreading the interaction forces from the solid mesh to the fluid grid. 
Boffi et al. \cite{Boffi2003} introduced natural interpolation operators between fluid and structure 
discrete spaces. 
Baaijens et al. \cite{baaijens2001fictitious} proposed the mortar element method for imposing a 
velocity continuity on 
the FSI interface with the use of Lagrangian multipliers. This approach was generalized by Hesch et 
al. \cite{hesch2014mortar} for 
enforcing the velocity constraint over the entire overlapping region between fluid and solid domain. A 
very similar approach based on Nitsche's Method \cite{nitsche1971} was proposed by Kamensky et al. \cite{kamensky2014variational} with the difference of restricting the coupling to the structure boundary.

In this work, we describe a novel FSI formulation based on the IBM. We employ a finite difference 
method for discretizing the incompressible flow and couple it with a finite element method for the
full elastodynamics equations of the structural 
problem by using an $L^2$-projection approach for handling the
interface conditions \cite{krause2016parallel}, namely the velocity continuity and force exchange. 
The framework is capable of coping
with general constitutive characteristics (including anisotropic materials) and complex flow 
configurations.

The algorithmic framework allows for the transfer of discrete fields between unstructured and 
structured meshes, which can be arbitrarily distributed among processors. This ensures convergence, efficiency, 
flexibility, load balancing, and accuracy without requiring a priori information on the relation between 
the different meshes.
Therefore, the approach introduced in this work is well suited for coupling already existing flow and structure solvers (legacy solvers).

The main novelties of the proposed method may be summarized as follows:
\begin{enumerate}

\item The transfer of data between the Eulerian finite difference grid of the fluid and the Lagrangian 
finite element mesh of the structure is achieved by a fully variational approach,
which does not require the use of pointwise interpolation schemes as in the classical IBM. 

\item The solid motion is modelled  by solving the elastodynamics equation via a fully implicit time-integration scheme, whereas other implementations 
of IBM derive the motion of the solid structure from the fluid velocity field \cite{griffith2012hybrid} or describe it through simplified kinematic equations~\cite{sotiropoulos2009review}. 

\item The use of a high-order Navier--Stokes solver allows for direct numerical simulations (DNS)
of laminar, transitional and turbulent flows.

\end{enumerate}

For constructing the transfer operator a partition of unity is assigned to each point of the fluid grid 
\cite{melenk1996partition}, i.e. basis functions are attached to the fluid grid.
We further introduce a new analytical benchmark problem for the verification of the correct implementation of inertial forces.

The article is divided into five sections. Following this introduction, the fundamental equations 
governing the FSI problem are presented (Section \ref{PF}). The second part of Section \ref{PF} is 
dedicated to the description of the coupling 
strategy and provides details about the variational transfer. Section \ref{sec:coupling} describes the 
entire framework and illustrates the FSI algorithm with a flow chart. Numerical results for various benchmark problems are presented and 
analyzed in Section \ref{sec:results}. Finally, 
some concluding remarks are drawn in Section \ref{sec:conclusion}. \ref{sec:inertiabench} gives 
mathematical details of the inertial benchmark used in Section \ref{sec:results}.

\section{Solid, Fluid and Interaction Problem Formulations and Discretizations}
\label{PF}
This section provides an overview of the equations governing the FSI problem. We adopt
a Lagrangian specification of the immersed structure and a Eulerian 
specification of the fluid.

\subsection{Solid Dynamics Formulation}
\begin{figure}
\begin{center}
\includegraphics[scale=0.40]{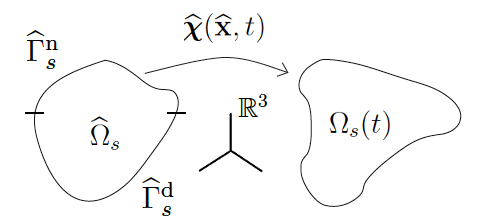}
\end{center}
\caption{Reference configuration (left) and current configuration (right) of a continuum body. Here 
the boundary $\partial\widehat{\Omega}_s$ is split into nonoverlapping Neumann $
\widehat{\Gamma}_{s}^{\rm n}$ and Dirichlet $
\widehat{\Gamma}_s^{\rm d}$ boundaries.}
\label{fig01}
\end{figure}

Let $\widehat{\Omega}_s\subset \mathbb{R}^3$ be a bounded Lipschitz domain. We refer to a body of mass 
undergoing a motion from the material (reference) configuration, $
\widehat{\Omega}_s$, to the current (spatial) configuration, ${\Omega}_s(t)$ (Figure~\ref{fig01}).
The material  position ${\bm{ \widehat{x}}}$  and the actual position  ${\bm{x}}$ are linked during the time interval $I:=[0\,,T]$ of interest by a one-to-one mapping, called {\it{motion}
} ${\widehat{\boldsymbol{\chi}}}: {\widehat{\Omega}_s }\times I \rightarrow  \mathbb{R}^d$, with spatial dimension $d\in \{1,2,3\}$ 
such that $\bm{x}={\widehat{\boldsymbol{\chi}}}({\bm {\widehat{x}}}, t)$, $\forall\,t\in I$.

\paragraph{Governing Equations}
As customary in solid dynamics, the solid subproblem is described in terms of the mapping ${\widehat{\boldsymbol{\chi}}}$.
The total Lagrangian specification of the elastodynamics balance equations for the solid domain is:
\begin{equation}
\label{momentum_equation}
  \widehat{\rho}_{s}\frac{\partial^{2}{\bm{\widehat{u}}}_{s}}{\partial{t}^{2}}
  -
  \widehat{{\nabla}}\cdot {\bm{\widehat{P}}}
  =
\bm{0}
  \quad \mathrm{on}
  \quad \widehat{\Omega}_{s}.
\end{equation}

Here  $\widehat{\rho}_{s}$ is the mass density per unit undeformed volume, $\widehat{\bm{u}}_{s}=
\widehat{{\bm u}}_{s}{(\widehat{{\bm x}},t})$ is the displacement field,  
$\widehat{{\bm P}}=\widehat{{\bm P}}({{\widehat{\bm x}},t})$ is the first Piola--Kirchhoff stress tensor, and
$\widehat{{\nabla}}\,\cdot$ is the divergence operator computed in the reference configuration. 

For a hyperelastic material the first Piola--Kirchhoff stress tensor 
$\widehat{{\bm {P}}}=\partial\Psi/\partial \bm{F}$ is related to the deformation 
through a constitutive equation derived from a given scalar energy function $\Psi$.
In this paper we will consider the Saint-Venant--Kirchhoff constitutive relation, ${\Psi}_{I}$, and the fiber-reinforced
model proposed by Holzapfel \cite{holzapfel2000new}, ${\Psi}_{II}$, which read as follows:
\begin{align}
{\Psi}_{I} & =\frac{\lambda}{2} [{\mathrm  {tr}}({\widehat{\bm{E}}
})]^2+\mu_s \mathrm  {tr}({\widehat{\bm{E}}^2}), \\
{\Psi}_{II} &=\mu_s (\bar{I}_1 - 3) + 
\frac{k_{11}}{2k_{21}}(\exp[{k_{21}(\bar{I}_{4,1} -1)^2}] -1)+  \frac{k_{12}}{2k_{22}}(\exp[{k_{22}(\bar{I}_{4,2} -1)^2}] - 1). \label{eq:Holzapfel}
\end{align}

\noindent Here ${\bm {\widehat{E} }} = ({\bm {\widehat{F}}}^T {\bm {\widehat{F}}} - \bm{I})/2$ is the Green-Lagrangian strain 
tensor, $\mathrm  {tr} (\cdot)$ is the trace operator, $\lambda_s$, $\mu_s$ and $k_{ij}$ are the constitutive parameters, and $\bar{I}_{1}$, $\bar{I}_{4,1}$ and $\bar{I}_{4,2}$, are modified invariants defined as 
\begin{equation}
\label{invariants}
\bar{I}_1=\widehat{J}^{-2/3}\mathrm  {tr}({\widehat{\bm{C}}})\qquad \bar{I}_{4,1}=\widehat{J}^{-2/3}
\bm{g_{0,1}}\cdot \widehat{\bm C} \bm{g_{0,1}}\qquad \bar{I}_{4,2}=\widehat{J}^{-2/3}\bm{g_{0,2}}\cdot 
\widehat{\bm C} \bm{g_{0,2}},
\end{equation}
with the unit vectors $\bm{g_{0,1}}$ and $\bm{g_{0,2}}$ denoting the fiber orientation, $\widehat{J}:=\det({\bm {\widehat{F}}})$ being
the determinant of the deformation gradient tensor and $\widehat{\bm {C}}:={\bm 
{\widehat{F}}}^T {\bm {\widehat{F}}}$ being the right Cauchy--Green strain tensor.

In order to fulfill the incompressibility condition, the penalty technique is employed. In this 
method, a volumetric energy term $\Psi_{V}(J) = 1/2 \kappa (J-1)^2$ is added to the expression of the 
strain energy function $\Psi$ with $\kappa$ representing the penalty coefficient.

Equation $\eqref{momentum_equation}$ must be supplied with initial conditions for the displacement field and the 
velocity field:
\begin{eqnarray}
\label{initial_conditions}
\nonumber
{\bm{\widehat{u}}}_{s}(\cdot,0)&=&{\bm{\widehat{u}}}_{s0}\quad\mbox{on}\, \widehat{\Omega}_s,\\
\frac{\partial{\bm{\widehat{u}}}_{s}(\cdot,0)}{\partial{t}}&=&{\bm{\widehat{v}}}_{s0}\quad\mbox{on}\, \widehat{\Omega}_s,
\end{eqnarray}
where ${\bm{\widehat{u}}}_{s0}$ and ${\bm{\widehat{v}}}_{s0}$ are given initial data, and with suitable boundary conditions. After splitting the boundary $\partial\widehat{\Omega}_s$ into the Neumann 
$\widehat{\Gamma}_{s}^{\rm n}$ and the Dirichlet $\widehat{\Gamma}_s^{\rm d}$ nonoverlapping parts (Figure~\ref{fig01}), the following boundary conditions are considered:
\begin{eqnarray}
\label{boundary_conditions}
\nonumber
{\bm{\widehat{u}}}_{s}&=&\bm{\widehat{b}}\quad\mbox{on}\,\widehat{\Gamma}_s^{\rm d},\\
{\bm{\widehat{P}}}\cdot \widehat{\bm{n}}_s&=&\bm{0}\quad\mbox{on}\,\widehat{\Gamma}_s^{\rm n},
\end{eqnarray}
where $\widehat{\bm{n}}_s$ is the outward normal and $\bm{\widehat{b}}$ is a prescribed boundary datum.
\paragraph{Weak Formulation}
Introducing the space of admissible test functions as
\begin{align}
\nonumber
\widehat{\bm{V}}_s&=\{\boldsymbol{\phi}_s \in \bm{H}^1(\widehat{\Omega}_s): \,\boldsymbol{\phi}_s|_{\widehat{\Gamma}_{s}^{\rm d}}=\bm 0 \},
\end{align}
where $\bm{H}^1 (\widehat{\Omega}_s)$ is the Sobolev space of weakly differentiable functions, 
the weak formulation of the elastodynamics balance 
equations \eqref{momentum_equation} reads:
 \begin{equation}
 \label{weak_form}
\int_{\widehat{\Omega}_s} \widehat{\rho}_{s0}\frac{\partial^2 {\widehat{\bm{u}}}_s}{\partial t^2}\cdot {\boldsymbol{\phi}_s} \,d\widehat{V}+ \int_{\widehat{\Omega}_s} \widehat{\bm{P}}:\widehat{\nabla}{\boldsymbol{\phi}_s} \,d\widehat{V} =\bm{0} \quad\forall \bm{\phi}_s\in \widehat{\bm{V}}_s.
\end{equation}

\paragraph{Spatial Discretization}  
We assume that the solid domain $\widehat{\Omega}_s$ can be approximated by a discrete domain $
\widehat{\Omega}_s^h$ and the associated mesh $\widehat{T}_s^h = \{\widehat{E}_s\subseteq\widehat{\Omega}
_s^h | \bigcup \widehat{E}_s = \widehat{\Omega}_s^h \}$, where its elements $\widehat{E}_s$ form a partition; hence for $E_s^1,\,E_s^2\subseteq\widehat{T}_s^h$ 
and $E_s^1\neq E_s^2$, then $E_s^1\cap E_s^2=\emptyset$. 

For the spatial discretization, we consider first-order finite elements for which the corresponding function space is defined as
\begin{equation}
\nonumber
\widehat{X}_s^h(\widehat{T}_s^h)=\{\phi_s^h \in C^0(\widehat{\Omega}_s^h) , \phi_{s}^h|_{\widehat{E}_s} \in {\mathbb{P}}_1\,\forall {\widehat{E}}_s 
\in \widehat{T}_s^h \},
\end{equation}
where $\mathbb{P}_1$  is the space of linear polynomials defined on each element ${\widehat{E}}_s\in \widehat{T}_s^h$.

Hence, the Galerkin formulation of the solid subproblem $\eqref{weak_form}$ reads:
\begin{equation}
\label{weak_form_solid}
\int_{\widehat{\Omega}_s}\widehat{\rho}_s\frac{\partial^2 \widehat{\bm{u}}^h_s}{\partial t^2}\cdot \bm{\phi}_s^h\,d\widehat{V} +\int_{\widehat{\Omega}_s}\widehat{ \bm{P}}(\widehat{\bm{u}}^h_{s}) : \widehat{\nabla} \bm{\phi}_{s}^h\,d\widehat{V}=\bm{0}\quad \forall \bm{\phi}_s^h\in \widehat{\bm{V}}_s^h.
\end{equation}

Let $\{\bm{N}_{s,i}^h\}_{i\in J_s}$ be the Lagrangian basis of the space 
$\widehat{\bm{V}}_s^h:= \widehat{\bm{V}}_s\cap\widehat{\bm{X}}_s^h$ where $J_s\subset\mathbb{N}$ is an index set, 
then the problem $\eqref{weak_form_solid}$ can be written as:
\begin{equation}
\label{spaceSP}
\widehat{\rho}_{s}\mathbf{M}\dfrac{\partial^2 \widehat{\mathbf{u}}_s}{\partial t^2}+\mathbf{K}
(\widehat{\mathbf{u}}_s) = \mathbf{0}.
\end{equation}
where $\widehat{\mathbf{u}}_s=[\widehat{\rm{u}}_{s,i} ]$ is the vector of the unknowns of the problem,  $\mathbf{M}$ is the mass matrix and $\mathbf{K}$ is the vector of nonlinear internal forces defined as follows:
\begin{align}
\nonumber
\mathbf{M}_{ij} &= \int_{\widehat{\Omega}_s} \bm{N}^h_{s,j}\cdot \bm{N}^h_{s,i}\,d\widehat{V},\\
\nonumber
\mathbf{K}(\widehat{\mathbf{u}}_s)_i & = \int_{\widehat{\Omega}_s}\widehat{ \bm{P}}(\widehat{\mathbf{u}}^h
_{s}) :  \widehat{\nabla}\bm{N}^h_{s,i}\,d\widehat{V}.
\end{align}

\paragraph{Time Discretization} 

The Newmark scheme~\cite{chung1993time,erlicher2002analysis} is adopted for the temporal discretization of the 
solid subproblem. Hence the discretized equation of motion $\eqref{momentum_equation}$ for a given discrete time step $n$ reads:

\begin{equation}
\label{Newmark}
{\widehat{\rho}_{s}}\mathbf{M}\frac{\widehat{{\mathbf{u}}}_s^{n+1}}{\Delta t^2}  ~+~ \beta\mathbf{K}(\widehat{\mathbf{u}}_s^{n+1}) 
~=~ \beta{\mathbf{F}}^{n} ,
\end{equation}
with 
\begin{align*}
{\mathbf{F}}^{n} :&=  \widehat{\rho}_{s}\mathbf{M}
\frac{\widehat{{\mathbf{u}}}_s^{n}}{\Delta t^2} +\widehat{\rho}_{s} \mathbf{M}
\frac{\widehat{{\mathbf{v}}}_s^{n}}{\Delta t}+\widehat{\rho}_s\mathbf{M} \frac{(1-2\beta)}{2}\widehat{\bf{a}}_{s}^{n}.
\end{align*}
Equation \eqref{Newmark} together with the following approximations:
\begin{eqnarray}
\widehat{\bf{u}}_{s}^{n+1}&=&\widehat{\mathbf{u}}^{n}_s+\Delta t \widehat{\mathbf{v}}^{n}_s 
+\frac{\Delta t^2}{2} \left(2\beta\widehat{\mathbf{a}}_{s}^{n+1} +(1- 2\beta) \widehat{\mathbf{a}}_{s}^{n}\right)\\
\widehat{{\mathbf{v}}}_{s}^{n+1}&=&\widehat{{\mathbf{v}}}_{s}^{n} + \begin{matrix}{\Delta t}\end{matrix}~((1-
\gamma)\widehat{\bf{a}}_{s}^{n} +\gamma\widehat{{\mathbf{a}}}_{s}^{n+1})
\end{eqnarray}
defines the Newmark scheme. Here $\widehat{{\mathbf{a}}}_{s}={\partial^2 \widehat{\mathbf{u}}_s}/{\partial t^2}$  and $\widehat{{\mathbf{v}}}_{s}={\partial \widehat{\mathbf{u}}_s}/{\partial t}$ denote the acceleration and velocity fields of the structure, respectively; $\beta$ and $\gamma$ are real parameters used to control the amplification of 
the high frequency modes which are not of interest. In the numerical experiments presented in Section 4,  we adopt the following set of parameters: $\beta=0.25$ and $\gamma=0.50$.

\paragraph{Implementation}The solid subproblem is implemented in the finite element framework 
MOOSE (\url{https://www.mooseframework.org}). A Newton method is used for solving the solid subproblem, whereas the 
MUltifrontal Massively Parallel Sparse direct Solver (MUMPS) is employed for solving the associated linear system. 

\subsection{Fluid Dynamics Formulation}
\label{FF}
The fluid dynamics subproblem is formulated in an Eulerian specification
where a bounded domain $\Omega_f$ is 
considered.
\paragraph{Governing Equations}
\label{ssec:goveq}
In the domain $\Omega_f$,  the Navier--Stokes equations for 
incompressible flow in nondimensional form are
\begin{subequations}
    \begin{equation}
    \label{momentum_equation_fluid}
        \frac{\partial {\widetilde{\bm{v}}}_f}{\partial \widetilde{t}} 	
        + \left({\widetilde{\bm{v}}}_f\cdot\widetilde{\nabla}\right){\widetilde{\bm{v}}}_f  
        + \widetilde{\nabla} \widetilde{p}_f - \frac{1}{\mathrm{Re}}\widetilde{\Delta}
        \widetilde{\bm{v}}_f 
        =\widetilde{\bm{f}},
    \end{equation}
    \begin{equation}
    \label{continuity_equation_fluid}
        \widetilde{\nabla}\cdot\widetilde{\bm{v}}_f = 0,
    \end{equation}
    \label{eq:NSE}
\end{subequations}
where the dimensional quantities have been 
nondimensionalized according to:
\begin{center}
	\begin{tabular}{ccc}
		${\bm{v}}_f = \widetilde{\bm{v}}_f U_\mathrm{ref}$, &
		${\bm{x}}    = \widetilde{\bm{x}} L_\mathrm{ref}$,    &
		${t} 				 = \widetilde{t} L_\mathrm{ref} / U_\mathrm{ref}$, \\
		${p}_f          = \widetilde{p}_f \rho_f U_\mathrm{ref}^2$,  &
		${\bm{f}}     = \widetilde{\bm{f}} \rho_f U_\mathrm{ref}^2 / L_\mathrm{ref}$, &
		$\mathrm{Re}   = \rho_f U_\mathrm{ref} L_\mathrm{ref} / \mu_f$.
	\end{tabular}
\end{center}
Here, $\bm{v}_f$ is the fluid velocity vector field, $\bm{x}$ is the coordinate vector with
components $x_{1,2,3}$, $p_f$ is the fluid pressure, $\bm{f}$ is an external force density; $\rho_f
$, $\mu_f$, $L_\mathrm{ref}$ and $U_\mathrm{ref}$ are the fluid density, dynamic 
viscosity, reference length and reference velocity, respectively; $\mathrm{Re}:=\rho_f U_
\mathrm{ref} L_\mathrm{ref}/\mu_f$ is the Reynolds number.

We introduce the notations $\mathcal{D}$ for the divergence operator, 
$\mathcal{L}\widetilde{\bm{v}}_f$ for the linear viscous term, 
$\mathcal{G}\widetilde p_f$
for the pressure gradient and $\mathcal{N}$ for all other terms in Equation \eqref{momentum_equation_fluid} except 
the temporal derivative.
As such the Navier--Stokes equations can be written in matrix operator form as:
\begin{equation}  
	\dfrac{\partial}{\partial \widetilde{t}}
    \begin{bmatrix}
    \widetilde{\bm{v}}_f\\[0.3em]
      0 \\[0.3em]
      \end{bmatrix} +
     \begin{bmatrix}
     -\mathcal{L} &  \mathcal{G} \\[0.3em]      
     \mathcal{D}  & 0            \\[0.3em]
     \end{bmatrix}
     \begin{bmatrix}
      \widetilde{\bm{v}}_f \\[0.3em]
      \widetilde p_{f} \\[0.3em]
      \end{bmatrix}
      =
      \begin{bmatrix}
     \mathcal{N}(\widetilde{\bm{v}}_f,\,\widetilde{\bm{f}})  \\[0.3em]
      0         \\[0.3em]
     \end{bmatrix}.
\end{equation}

\paragraph{Time Discretization}
\label{TimeFM}
Although our framework allows also for a semi-implicit time integration scheme, an explicit 
low-storage third-order three-stage Runge--Kutta method~\cite{wray1986very} is adopted for the time discretization of the fluid subproblem. 
Because the source force term in the FSI formulation (see Section \ref{ssec:FSI}) imposes a time 
step restriction which is more stringent than the CFL-like (Courant--Friedrichs--Lewy) stability 
condition arising from the viscous term \cite{stockie1999analysis}, the computational cost of an 
implicit treatment of the viscous term cannot be justified by a larger time step size.

The use of the explicit time integrator leads to a coupled system of
linear equations for the velocity $\widetilde{\bm{v}}_{f}^{(m)}$ and the pressure 
$\widetilde p_{f}^{(m)}$ at the subtime 
step $m=\{1,2,3\}$ which reads:
\begin{equation}
\begin{bmatrix}
     \mathcal{I} & c_m\Delta \widetilde t \mathcal{G} \\[0.3em]      
     \mathcal{D}  & 0            \\[0.3em]
     \end{bmatrix}
     \begin{bmatrix}
      \widetilde{\bm{v}}_f^{(m)} \\[0.3em]
      \widetilde p_{f}^{(m)} \\[0.3em]
      \end{bmatrix}
      =
      \begin{bmatrix}
      q(\widetilde{\bm{v}}_f^{(m-1)}, \widetilde{\bm{v}}_f^{(m-2)},
      \widetilde{\bm{f}}) \\[0.3em]
      \bm{0}         \\[0.3em]
\end{bmatrix},
\label{eq:timesystem}
\end{equation}
where $\mathcal{I}$ is an identity matrix, $q$ contains the right-hand side arising from 
the low storage Runge--Kutta scheme and $c_m$ is the Runge--Kutta stage coefficient.
\begin{figure}
\centering
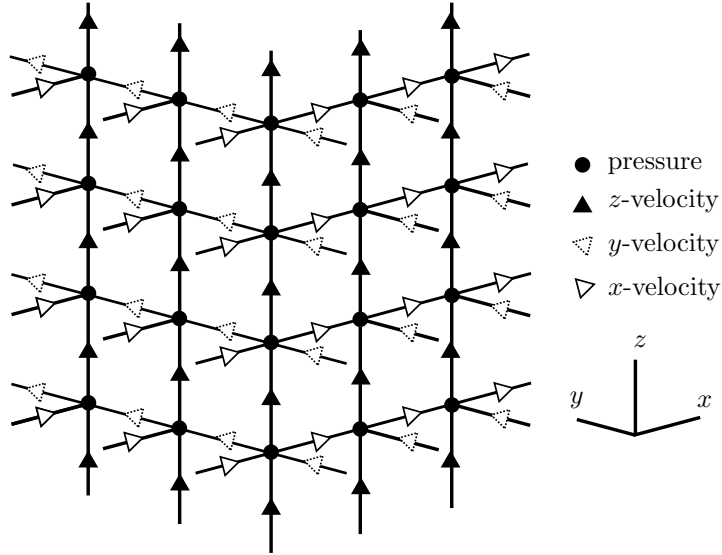
\caption{Three-dimensional staggered grid adopted for the fluid subproblem.}
\label{fig02}
\end{figure}

\paragraph{Spatial Discretization}
We use finite differences of high convergence order (sixth-order) on a rectilinear structured grid for the spatial discretization of Equation~$\eqref{eq:timesystem}$ \cite{henniger2010high}. This leads to a linear system of equations of the form:	
\begin{equation}
\label{spaceNS}
 \begin{bmatrix}
     \mathbf{J} & \mathbf{G} \\[0.3em]      
     \mathbf{D}  & \mathbf{0}            \\[0.3em]
     \end{bmatrix}
     \begin{bmatrix}
      \widetilde{\mathbf{v}}_{f} \\[0.3em]
      \widetilde{\mathbf{p}}_{f} \\[0.3em]
      \end{bmatrix}
      =
      \begin{bmatrix}
      \mathbf{q}\\[0.3em]
      \mathbf{0}         \\[0.3em]
     \end{bmatrix}
\end{equation}
Here the matrices $\mathbf{D}$ and $\mathbf{G}$ are the spatial discretization of the operators 
$\mathcal D$ and $\mathcal{G}$, respectively,  $\mathbf{q}$ is the discrete representation of the right-hand side $q$
and the discrete identity matrix $\mathbf{J}$ also contains the values of the velocity boundary conditions.
We work with four subgrids, one for each velocity component and one for the pressure (Figure~\ref{fig02}). The 
momentum equations are solved on the
respective velocity grids, which implies that the discrete operator $\mathbf{G}$ requires the evaluation of
the first derivatives of the velocity grids $1, 2, 3$ from values on the grid $0$. The continuity equation is 
satisfied on the pressure grid, i.e. the discrete operator $\mathbf{D}$ computes the first derivative on 
the pressure grid $0$ from the values of the grids $1,2,3$.

We can derive an equation for the pressure by forming the Schur complement of Equation~\eqref{spaceNS}:
\begin{equation}
\mathbf{D}\mathbf{J}^{-1} \mathbf{G} \widetilde{\mathbf{p}}_f= \mathbf{D}\mathbf{J}^{-1} \mathbf{q}.
\label{eq:PoissonDisc}
\end{equation}

The Poisson problem \eqref{eq:PoissonDisc} is solved with the iterative Krylov subspace method 
BiCGstab with right 
preconditioning by a \emph{V}-cycle geometric multigrid preconditioner of Gauss--Seidel type
\,\cite{henniger2010high}. 
To aid convergence we compute the left
null-space of the pressure operator and project it onto the column space of the operator as described in 
\cite{henniger2010high,simens2009high}.
\paragraph{Implementation}The described numerical approach is implemented in the 
Navier--Stokes solver IMPACT which is thoroughly validated and has been used for several complex flow configurations
~\cite{burns2015sediment, henniger2010direct, john2014stabilisation}. 
More details on this solver can be found in \cite{henniger2010high}.

\subsection{Fluid-Structure Coupling}
\label{ssec:FSI}
\begin{figure}
\begin{center}
\includegraphics[width=\textwidth]{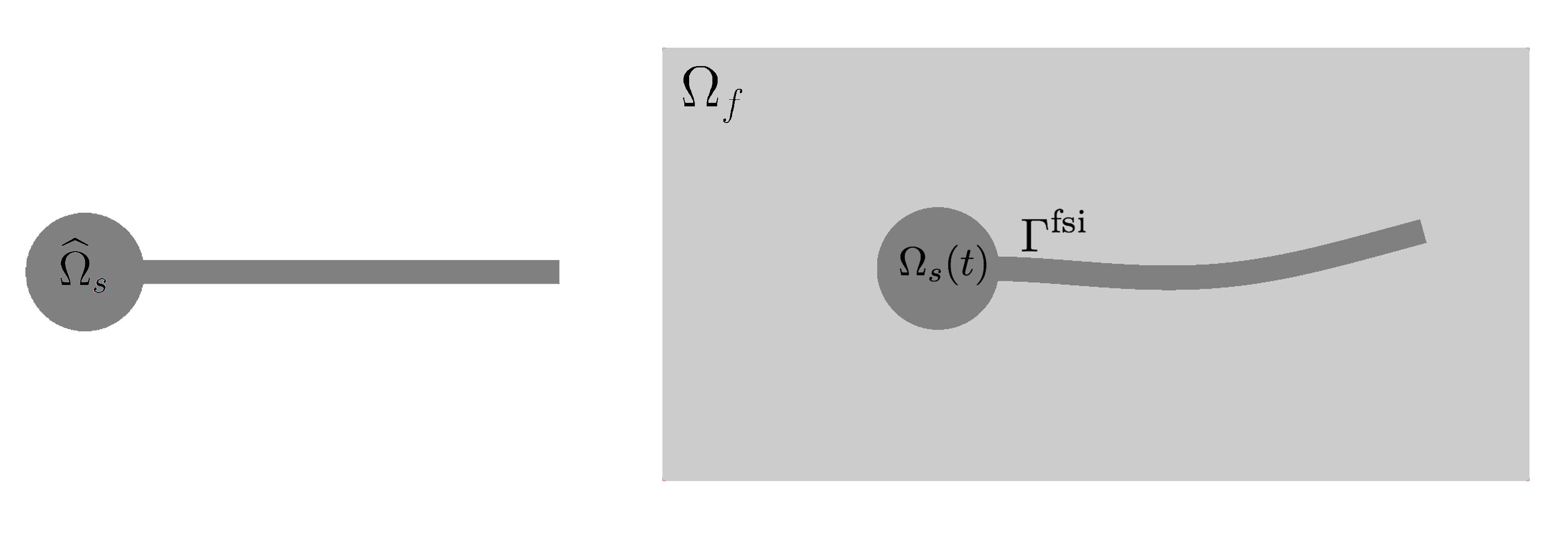}
\end{center}
\caption{(Left) Reference configuration of the solid domain (${\widehat{\Omega}_s}$). (Right)  
Eulerian representation of the fluid domain $({\Omega}_f)$ in which the current configuration of the 
structure (${\Omega}_{s}(t)$) is immersed. $\Gamma^\text{fsi}$ represents the boundary of fluid-structure interaction.} 
\label{fig03}
\end{figure}

The coupling between the discretizations of the fluid and the solid subproblems is established by enforcing congruent velocities at 
the interface $\Gamma^{\rm{fsi}}$ between fluid and structure (Figure~\ref{fig03}) and by adding a force density term to the 
Navier--Stokes equations to account for the immersed solid structure.
The strong formulation of the FSI problem reads as follows:
\begin{subequations}
\label{FSI_strong_form}
\begin{align}
\widehat{\rho}_{s}\frac{\partial^{2}{\bm{\widehat{u}}}_{s}}{\partial{t}^{2}} - \widehat{{\nabla}}\cdot {\bm{\widehat{P}}} =& \,{\bm {0}} 
\quad\quad\quad\quad \mathrm{in}
\quad\widehat{\Omega}_{s} \label{eq:FSI_strong_form1}\\
\frac{\partial \widetilde{\bm{v}}_f}{\partial \widetilde{t}} + \left(\widetilde{\bm{v}}_f\cdot\widetilde{\nabla}\right)\widetilde{\bm{v}}_f  + \widetilde\nabla \widetilde p_f - \frac{1}
{\mathrm{Re}}\widetilde\Delta\widetilde{\bm{v}}_f =&\,\widetilde{\bm{f}}_{\text{fsi}}\quad\quad\,\,\,\,\,\,\,\mathrm{in}
\quad {\Omega}_{f} \label{eq:NSEa}\\
\widetilde\nabla\cdot\widetilde{\bm{v}}_f =&\,0\,\,\,\quad\,\,\quad\quad\,\mathrm{in}\quad {\Omega}_{f}\label{eq:NSEb}\\
\frac{\partial \bm{u}_s}{\partial t}=&\,{\bm{v}}_f\,\,\quad\quad\quad\,
\mathrm{on}\quad\Gamma^{\text{fsi}}\\
\bm{v}_f=&\,{\bm{v}_b} \,\,\,\quad\quad\quad\,\mathrm{on}\quad{\partial\Omega_f} \
\end{align}
\end{subequations}
where $\partial\Omega_f$ is the boundary of the fluid domain and ${\bm{f}}_{\mathrm{fsi}}   = \widetilde{\bm{f}}_{\mathrm{fsi}} \rho_f U_\mathrm{ref}^2/L_\mathrm{ref}$ 
 is the reaction force density generated by the immersed solid. It is computed as:
\begin{equation}
\label{reaction_f}
 \int_{\widehat{\Omega}_s} {\bm{f}}_{\mathrm{fsi}}  \cdot \bm{\phi}_s\,d\widehat{V}=  
 \int_{\widehat{\Omega}_s}\widehat{\rho}_{s}\frac{\partial{^2}\widehat{\bm{u}}_s}{\partial t^2}\cdot\bm{\phi}_s \,d\widehat{V} +
 \int_{\widehat{\Omega}_s} \widehat{\bm{P}}:{\widehat \nabla}\bm{\phi}_s \,d\widehat{V}.
\end{equation}

Equations $\eqref{FSI_strong_form}$ are supplied with initial conditions for the displacement and
velocity field of the solid structure and for the velocity field of the fluid domain:
\begin{eqnarray}
\label{initial_conditions_FSI}
{\bm{\widehat{u}}}_{s}(\bm{x},0)&=&{\bm{\widehat{u}}}_{s}^{0} \quad\,\,\,\,\mbox{in}\quad\widehat{\Omega}_s\\
\frac{\partial{\bm{\widehat{u}}}_{s}}{\partial{t}}(\bm{x},0)&=&\dfrac{\partial{\bm{\widehat{u}}}_{s}^{0}}{\partial{t}} \quad \mbox{in}\quad\widehat{\Omega}_s\\
{\bm{v}}_{f}(\bm{x},0)&=&{\bm{v}}_{f}^{0}\quad\quad \mbox{in}\quad {\Omega}_f
\end{eqnarray}

\begin{figure}
\centering
\includegraphics[scale=0.09]{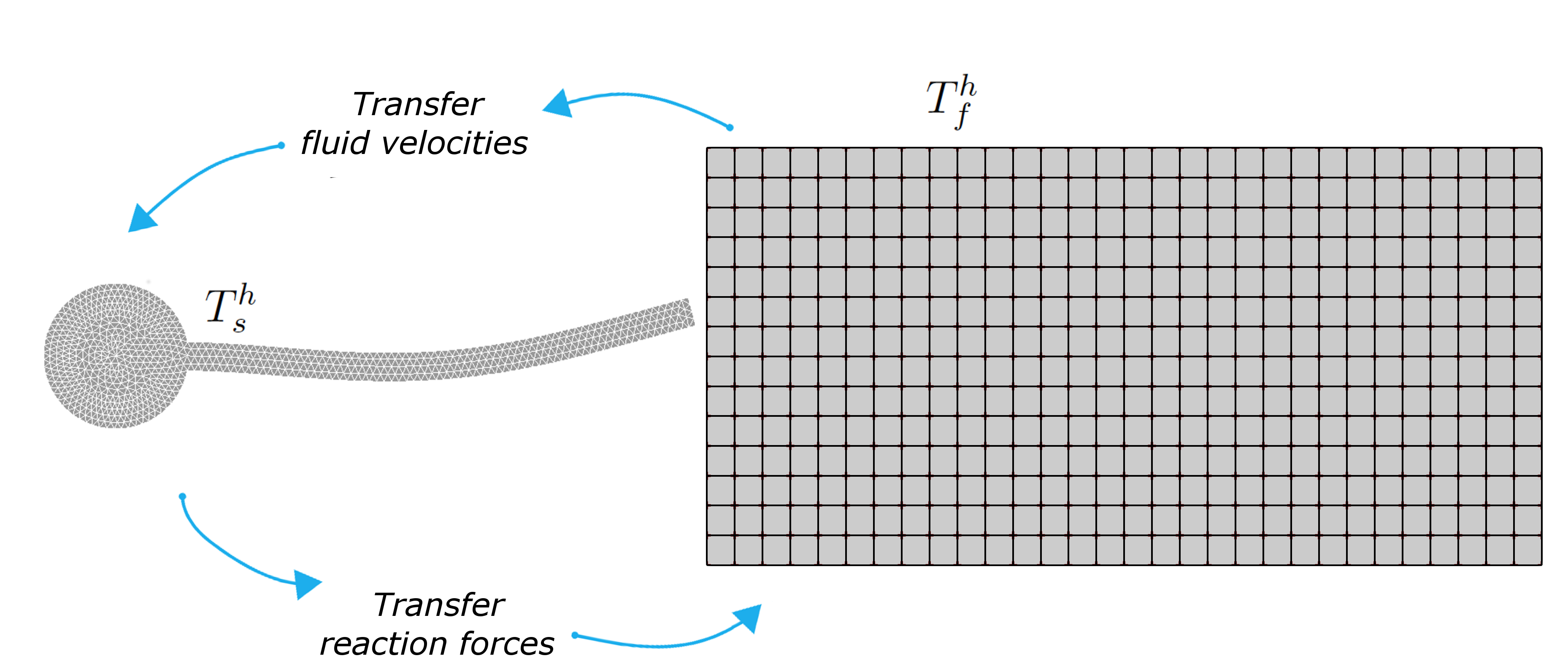}
\caption{Schematic representation of the coupling strategy. Deformed solid mesh, $T_s^h$, on the left and fluid grid, $T_f^h$, on the right. }
\label{fig:schematicFSI}
\end{figure}

\paragraph{Variational Transfer: the $L^2$-Projection Approach}
The coupling between fluid and structure requires the transfer of the velocities ${\bm{v}}_{f}$ and the force density $\bm{f}_{\mathrm{fsi}}$ from the Eulerian fluid grid to the Lagrangian solid mesh and vice versa (Figure~\ref{fig:schematicFSI}).

In the classical IBM, the coupling between the two types of variables involves a smoothed approximation of the Dirac-delta function.
It is well known that such an approach can suffer from poor volume conservation [2, 3]. This manifests itself as an 
apparent fluid \textit{leak} at fluid-structure interfaces, which occurs even though the Lagrangian structure moves at 
the local fluid velocity. This leaking can be observed as a numerical artifact that appears when the fluid 
element size is much smaller than that of the structure.  A heuristic estimation of mesh ratio of two is recommended to 
prevent leaking \cite{peskin1993improved}.

In this work, the finite difference discretization of the fluid dynamics subproblem and the finite 
element 
discretization  of the solid dynamics subproblem are coupled by means of $L^2$-projections. This 
coupling approach allows for 
the transfer of discrete fields between unstructured and structured discretized domains in a 
transparent, efficient, and flexible way.

The use of the $L^2$-projection approach requires to attach Lagrangian basis functions to the finite difference grid~\cite{fackeldey2011coupling}, and to define the corresponding auxiliary finite element space as  $\mathbf{V}_f ^h= \mathbf{V}_f^h(T_{f}^{h}) \subset [H_0^1(\Omega_{f})]^d$  where $T_{f}^h$ indicates the fluid 
grid (Figure~\ref{fig:schematicFSI}). Further, we introduce a suitable discrete space of Lagrangian multipliers $\mathbf{M}_{\mathrm{fsi}}^h(T_s^h)$, where $T_s^h$ represents the current configuration of the solid mesh. 

We introduce the FSI projection operator $\Pi: {V}_f^h \rightarrow {{V}_s^h} $ for the transfer of 
the discrete velocity field from the fluid grid to the solid mesh. For each scalar component of the velocity ${\mathrm{v}}_{f}^{h} \in {V}_f^h$ we want to find $\mathrm{w}_{s}^{h} =\Pi({\mathrm{v}}_{f}^{h}) \in{V}_s^h$, such that:
\begin{equation}
\label{discrete_operator}
\int_{I_h}({\mathrm{v}}_{f}^{h}-\Pi({\mathrm{v}}_{f}^{h,i}))\lambda_\mathrm{fsi}^{h}\,dV=
\int_{I_h}({\mathrm{v}}_{f}^{h}-\mathrm{w}_{s}^{h})\lambda_\mathrm{fsi}^{h}=0\,dV\quad \quad \forall\,\,\, 
\lambda_{\mathrm{fsi}}\in M_{\mathrm{fsi}}^{h}.
\end{equation}
where $I_h$ denotes the overlapping region between fluid grid and structure mesh, $I_h:=T_s^h\cap\,T_f^h$, here coinciding with the solid mesh $T_s^h$.
Let $\{N_{f,i}^h\}_{i\in J_{f}}$ and $\{N_{\mathrm{fsi},i}^h\}_{i\in J_{\mathrm{fsi}}}$ be the Lagrangian basis functions of the 
spaces $\mathbf{V}_f^h$ and $\mathbf{M}_{\mathrm{fsi}}^h$, respectively, with $J_f \in \mathbb{N} $ and
$J_\mathrm{fsi} \in \mathbb{N} $ suitable index sets,  then we get the so-called mortar integrals:
\begin{equation}
\mathbf{B}_{ij}= \int_{I_h}N^h_{f,j}N^h_{\mathrm{fsi},i}\,dV \quad\quad \mathbf{S}_{ij} = \int_{I_h}N^h_{s,j}
N^h_{\mathrm{fsi},i}\,dV.
\end{equation}
Equation $\eqref{discrete_operator}$ can be written in algebraic form:
\begin{equation}
\mathbf{B}{\mathbf{v}}_f=\mathbf{S}\mathbf{w}_s
\end{equation}
where $\mathbf{w}_s$ and ${\mathbf{v}}_f$ are vectors of coefficients entries $\mathrm{w}_{s}^{h,i}$ and ${\mathrm{v}}_f^{h,i}$.
In the present case $\mathbf{S}$ is square-shaped, thus one may compute the transfer operator $\mathbf{T}$ as:
\begin{equation}
\mathbf{w}_s=\mathbf{S}^{-1}\mathbf{B}{\mathbf{v}}_f=\mathbf{T}{\mathbf{v}}_f
\end{equation}
To reduce the computational cost required to compute the inverse of the matrix $\mathbf{S}$, dual basis 
functions may be adopted for the functional space $M^h_{\mathrm{fsi}}$. In this case the functional space is spanned 
by a set of functions which are biorthogonal to the basis functions of ${V}_s^h$ with respect to the $L^2$ inner product:
\begin{equation}
\label{biorth}
(N_{j,s}^{h}, N_{k,\mathrm{fsi}}^{h})_{L^2(I^h)}=\delta _{j.k}(N_{k,\mathrm{fsi}}^{h},1)_{L^2(I^h)}\quad\forall 
{i,j}
\end{equation}
The usage of the dual basis functions corresponds to replacing the standard $L^2$-projection with the local approximation (Equation~\eqref{biorth}) which we call `pseudo'  $L^2$-projection. This choice allows for a more efficient evaluation of the transfer operator $\mathbf{T}$ since the matrix  $
\mathbf{S}$ becomes diagonal.
Finally, we use the transpose operator $ \mathbf{T}^{T}$ to transfer the reaction forces from the solid mesh to the fluid grid:
\begin{equation}
\widetilde{\mathbf{f}}_{\mathrm{fsi}}=L_\mathrm{ref}/(\rho_f U_\mathrm{ref}^2) \mathbf{T}^{T}{\bf{f}}_{\mathrm{fsi}}.
\end{equation}
Here $\widetilde{\mathbf{f}}_{\mathrm{fsi}}$ is obtained by making nondimensional the $L^2$-projection of the vector ${\bf{f}}_{\mathrm{fsi}}\in \mathbf{V}_s^h(T_s^h)$  corresponding to the reaction force  ${\bm{f}}_{\mathrm{fsi}}$ defined in Equation~\eqref{reaction_f}.

The transfer operator is assembled as follows: (I) detect the overlapping region by means of a tree-search 
algorithm, (II) generate the quadrature points for integrating in the intersecting region, (III)  compute the local 
element-wise contributions for the operators $\mathbf{B}$ and $\mathbf{S}$ by means of numerical quadrature rules 
and (IV) finally assemble the two mortar matrices. The current implementation of the procedure described in~\cite{krause2016parallel} generates quadrature points exclusively for piecewise affine meshes. The necessity of transferring data in the current configuration of the solid mesh $T_s^h$ motivates the choice of $\mathbb P_1$ elements for the discretization of solid subproblem. 

\begin{remark}
In the classical IBM, the evaluation of the dynamic terms in the solid structure in Equation~\eqref{eq:FSI_strong_form1} is carried out
using the difference between the solid and the fluid densities $\widehat{\rho}_s-\rho_f$~\cite{hesch2014mortar}.
Here, this difference in density is only applied to the boundary $\Gamma^{\rm{fsi}}$, because
we enforce the displacement from the flow field only on the boundary $\Gamma^{\rm{fsi}}$. Further, we choose not to eliminate
the fluid stress terms at the interface with the solid domains from the Navier--Stokes equations (Equation \eqref{eq:NSEa}), because
the fluid stresses are typically considered negligible compared to the solid stresses imposed on the interface $\Gamma^{\rm{fsi}}$
~\cite{hesch2014mortar}.
\end{remark} 

\section{Fluid-Structure Interaction algorithm}
\label{sec:coupling}
\begin{figure}
\begin{center}
\begin{tikzpicture}[thick,scale=0.75, every node/.style={transform shape}]
\centering
\node(start)[block step,
        	label={[label distance=0.1cm]330: $t=t^n$}
        ]{
			$\mathbf{v}_f^n$, ${\mathbf{f}}_{\mathrm{fsi}}^n$, $\widehat{\mathbf{u}}_s^n$
		 };
\node(picardinit)[block picardstep2,
		   anchor=north,
		   at={([yshift=-\vertSpacing]start.south)}
         ]{
         	$\mathbf{v}_{f,l=0}=\mathbf{v}_{f}^n$, ${\mathbf{f}}_{\mathrm{fsi},l=0}={\mathbf{f}}_{\mathrm{fsi}}^n$, 
		 	$\widehat{\mathbf{x}}_{s,l=0}^n=\widehat{\mathbf{x}}_s^n$, $\widehat{\mathbf{u}}_{s,l=0}^n=\widehat{\mathbf{u}}_s^n$
         };
\node(iterate)[block picard,
		   anchor=north,
		   at={([yshift=-\vertSpacing]picardinit.south)},
		   label={[label distance=0.2cm]0: $\text{with}\, l \in \mathbb{N}$}
		 ]{
		 };
\node(bcs)[block picardstep,
		   anchor=north,
		   at={([yshift=-\vertSpacing]iterate.south)},
		   label={[label distance=0.2cm]0: \bf{Step 1}}
         ]{
         	${\mathbf{w}}_{s,l}=\mathbf{T}{\mathbf{v}}_{f,l-1}$
         };
\node(structure)[block picardstep,
		   anchor=north,
		   at={([yshift=-\vertSpacing]bcs.south)},
		   label={[label distance=0.2cm]0: \bf{Step 2}}
         ]{
         $\mathbf{ \widehat{u}}^{\Gamma_{\rm{fsi}}}_{s,l} = \widehat{\mathbf{u}}_{s,0} + \Delta t\,\mathbf{w}_{s,l}$	
         };
\node(forcetransfer)[block picardstep,
		   anchor=north,
		   at={([yshift=-\vertSpacing]structure.south)},
		   label={[label distance=0.2cm]0: \bf{Step 3}}
         ]{
         	
	$[{\mathbf{f}}_{\mathrm{fsi},l},\widehat{\mathbf{u}}_{s,l}]=
	\widehat{S}\left(\mathbf{ \widehat{u}}^{\Gamma_{\rm{fsi}}}_{s,l}\right)$
         };
\node(impact)[block picardstep,
		   anchor=north,
		   at={([yshift=-\vertSpacing]forcetransfer.south)},
		   label={[label distance=0.2cm]0: \bf{Step 4}}
         ]{
        $\widetilde{\mathbf{f}}_{\mathrm{fsi}}=L_\mathrm{ref}/(\rho_f U_\mathrm{ref}^2) \mathbf{T}^{T}{\bf{f}}_{\mathrm{fsi}}$
         };
\node(veltransfer)[block picardstep,
		   anchor=north,
		   at={([yshift=-\vertSpacing]impact.south)},
		   label={[label distance=0.2cm]0: \bf{Step 5}}
         ]{
         $\mathbf{v}_{f,l}={F\left({\mathbf{v}}_{f,l},p_f\right)}$	
         };
\node(convergence)[block picardstep,
		   anchor=north,
		   at={([yshift=-\vertSpacing]veltransfer.south)},
		   label={[label distance=0.2cm]0: \bf{Step 6}}
         ]{
         $\text{Compute residual norms}$ $\text{and check for convergence}$	
         };
\node(termination)[block decision,
		   anchor=north,
		   at={([yshift=-\vertSpacing]convergence.south)}
         ]{
         	$\text{Converged?}$ 
         };
\node(end)[block step2,
		   anchor=north,
		   at={([yshift=-\vertSpacing]termination.south)},
		   label={[label distance=0.1cm]345: $t=t^{n+1}$}
		 ]{
			$\mathbf{\widehat{u}}_s^{n+1}=\mathbf{\widehat{u}}_{s,l}$,\\ 
			$\widetilde{\mathbf{f}}^{n+1}=\widetilde{\mathbf{f}}_{l}$,\\ 
			$\mathbf{v}_{f}^{n+1}\,\,=\mathbf{v}_{f,l}$
		 };

\begin{scope}[->]
  \draw (start) -- (picardinit);
  \draw (picardinit) -- (iterate);
  \draw (iterate) -- (bcs);
  \draw (bcs) -- (structure);
  \draw (structure) -- (forcetransfer);
  \draw (forcetransfer) -- (impact);
  \draw (impact) -- (veltransfer);
  \draw (veltransfer) -- (convergence);
  \draw (convergence) -- (termination);
  \draw (termination) -- node [near start, xshift=0.35cm] {yes} (end);
  \draw (termination) -- ++(-4,0) node [near start,  yshift=0.2cm] {no} |- (iterate);

\end{scope}

\end{tikzpicture}
\end{center}
\caption{Flow chart of the FSI algorithm. Here $\widehat{S}$ is the solid subproblem and $F$ is the fluid subproblem.}
\label{fig23}
\end{figure}
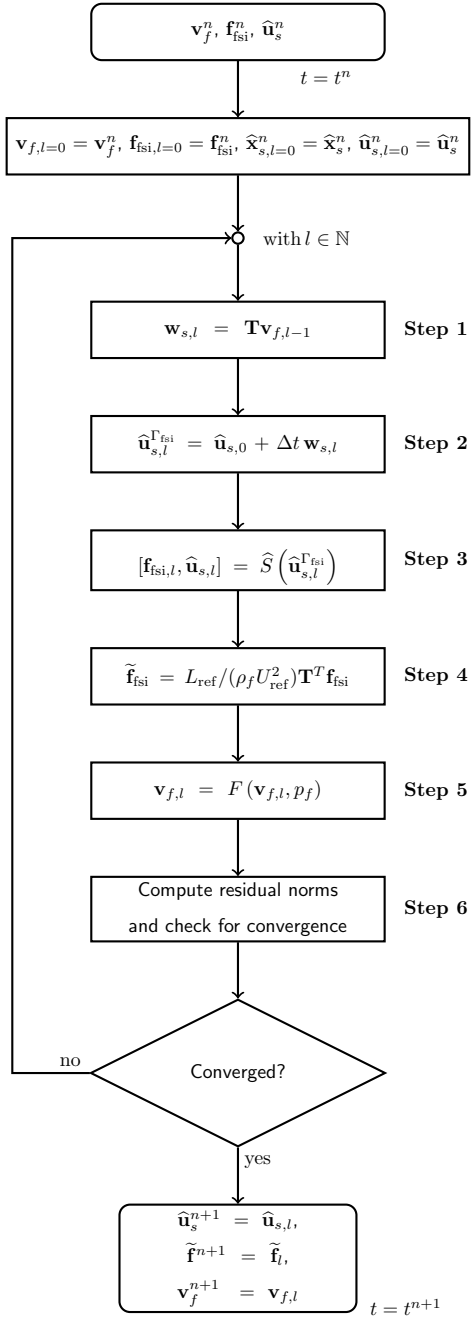

For solving the discretized FSI problem we adopt a segregated approach with a fixed 
point (Picard) iteration at each time step. To ensure numerical stability in time of the coupled nonlinear FSI system 
(Figure~\ref{fig23}), the system is solved to a sufficient spatial accuracy in each time step.

For a given time step $n$ and given a starting solution at the Picard iteration $p=l$ with $l\in\mathbb{N}$, the FSI 
algorithm determines the solution at the next time step $n + 1$ as follows:

\begin{enumerate}
\item [] {\bf{Step 1}}: Transfer the fluid velocity from the fluid grid to the current configuration of the solid mesh.
\end{enumerate}
\begin{equation}
\mathbf{w}_{s,l}=\mathbf{T}{\mathbf{v}}_{f,l-1}
\end{equation}
\begin{enumerate}
\item [] {\bf{Step 2}}: Compute the displacement field of the solid structure on the interface $\Gamma_{\rm{fsi}}$ and use it as a boundary condition for  
the elastodynamics equations $\eqref{spaceSP}$.
\begin{equation}
\mathbf{ \widehat{u}}_{s,l} = \widehat{\mathbf{u}}_{s,0} + \Delta t\,\mathbf{w}_{s,l}
\end{equation}
\end{enumerate}

\begin{enumerate}
\item [] {\bf{Step 3}}: Solve the elastodynamics equations $\eqref{spaceSP}$ and compute the reaction force $
{\mathbf{f}}_{\mathrm{fsi},l}$.
\end{enumerate}

\begin{enumerate}
\item [] {\bf{Step 4}}:  Transfer the reaction force ${\mathbf{f}}_{\mathrm {fsi},l}$ from the current configuration of the solid mesh
to the fluid grid.
\end{enumerate}

\begin{equation}
\widetilde{\mathbf{f}}_{\mathrm{fsi},l}=L_\mathrm{ref}/(\rho_f U_\mathrm{ref}^2)\mathbf{T}^{T}{\mathbf{f}}
_{\mathrm{fsi},l}
\end{equation}

\begin{enumerate}
\item [] {\bf{Step 5}}: Solve the Navier--Stokes equations \eqref{eq:NSEa},\eqref{eq:NSEb} by using the force $\widetilde{\mathbf{f}}_{\mathrm{fsi},l}$
as source term to get the new velocity value $\mathbf{v}_{f,l}$.
\end{enumerate}

\begin{enumerate}
\item [] {\bf{Step 6}}: Compute residual norms of the difference between the two latest available sets of FSI
force terms and compare them with a given threshold as follows:\\
\textit{Absolute convergence criterion}
\begin{equation}
\label{AC}
{\|{\mathbf{f}}_{\mathrm{fsi},l}-{\mathbf{f}}_{\mathrm{fsi},l-1}\|_{\infty}}<
\epsilon_{A}
\end{equation}
\textit{Relative convergence criterion}
\begin{equation}
\label{RC}
\frac{\|{\mathbf{f}}_{\mathrm{fsi},l}-{\mathbf{f}}_{\mathrm{fsi},l-1}\|_{\infty}}{\|{\mathbf{f}}_{0,\mathrm{fsi}}\|}<\epsilon_{R}
\end{equation}

Start a new Picard iteration if neither of these conditions are satisfied. Advance to a new time step if one of the two criteria is satisfied.

\end{enumerate}

\paragraph{Implementation}
The FSI algorithm is implemented in the finite element framework MOOSE and includes an interface with the library MOONoLith (\url{https://bitbucket.org/zulianp/par\_moonolith}) and the flow solver IMPACT.

\begin{figure}[htb]
\centering
\includegraphics[width=\textwidth]{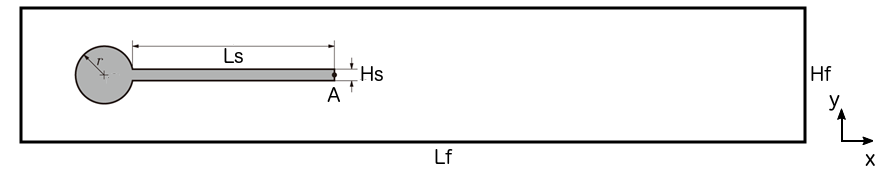}
\caption{Geometry of the Turek--Hron benchmark.}
\label{fig_bench_TH}
\end{figure}

\section{Numerical Results}
\label{sec:results}
A series of numerical simulations are presented in order to demonstrate the accuracy, robustness and flexibility of the 
developed computational framework. 
We present examples for moderately high Reynolds numbers. 
All computations have been performed on the \emph{Piz Daint} supercomputer at CSCS (Lugano, Switzerland), a hybrid Cray XC40/XC50 
system with a total of 5320 hybrid (GPU/CPU) compute nodes 
equipped with a 12-core 64-bit Intel Haswell CPU (Intel® Xeon® E5-2690 v3), an NVIDIA® Tesla® P100 with 64 GB of hybrid memory. 

\subsection{Turek--Hron FSI benchmark}
In this section, we present results for the Turek--Hron FSI benchmark 
\cite{turek2006proposal} of an
incompressible flow past an elastic solid structure.

The fluid domain (Figure~\ref{fig_bench_TH}) has a length of $L_f=3\,[\si{\meter}]$ and 
height $H_f = 0.41\,[\si{\meter}]$, whereas the 
immersed solid structure is composed of a  disk with radius  $r = 0.05\,[\si{\meter}]$ 
centered at $C = (0.2\, [\si{\meter}], 0.2\, [\si{\meter}])$ 
(measured from the left bottom corner of the channel)  and a tail consisting of a rectangular elastic beam of length 
$L_s = 0.35\,[\si{\meter}]$ and height 
$H_s = 0.02\,[\si{\meter}]$; its right bottom corner is positioned at $(0.6\, 
[\si{\meter}], 0.19\, [\si{\meter}])$,
and the left end is fixed to the circle. 
The fluid domain is discretized using a Cartesian grid with $769\times129$ 
grid points which is stretched to concentrate points around the structure.
The solid mesh consists of $3273$ linear finite elements ($\mathbb{P}_1$)  with $1791$ nodes. Bilinear finite elements ($\mathbb{Q}_1$) are used for the auxiliary function space $\mathbf{V}_f^h$ associated with the fluid subproblem for the assembly of the transfer operator $\mathbf{T}$.

Periodic boundary conditions are imposed at the inlet and outlet of the fluid channel together with no-slip 
boundary conditions on the top and the bottom. 
A parabolic velocity profile $\bm{v}_0(\bm{x},t)$ 
\begin{equation}
\bm{v}_0 = 1.5\,U_{\mathrm{ref}}\frac{y(H_f - y)}{H_f^2/4},
\end{equation}
is enforced upstream of the structure by adding a fringe forcing term~\cite{nordstrom1999fringe} to the right-hand side of the Navier--Stokes equations.  The fringe force acts in the region 
$x_\mathrm{start} < x < x_\mathrm{end}$ and is defined by the function $\lambda(x)$:
\begin{equation}
\lambda (x) = \hat \lambda \left[ 
S\left(
\frac{x - x_\mathrm{start}}{d_\mathrm{rise}}
\right)
-
S\left(
\frac{x - x_\mathrm{end}}{d_\mathrm{fall}} + 1
\right)
\right]\, ,
\end{equation}
and
\begin{equation}
S(x) = \begin{cases}
0\, , & x \leq 0\, ; \\
\left(1 + \exp\left(\frac{1}{1-x} + \frac{1}{x}\right)\right)^{-1} & 0 < x < 1\, ; \\
1\, , & x \geq 1\, , \\
\end{cases}
\end{equation}

with $x_\mathrm{start}  = 2.5$, $x_\mathrm{end} = 3.0$, $d_\mathrm{rise} = d_\mathrm{fall} = 0.025$ 
and $\hat\lambda=10$.
In this region we enforce the velocity profile $\bm{v}_0=U_\mathrm{ref}\widetilde{\bm{v}}_0$ and an appropriate fluid pressure increase by
\begin{equation}
\widetilde{\bm{f}}_\mathrm{fringe} = 
\lambda (\widetilde x)\left[(\widetilde{\bm{v}}_0 - \widetilde{\bm{v}}_f) + \frac{\widetilde {L}_f}{\hat\lambda(\widetilde {x}_\mathrm{end} - \widetilde {x}_\mathrm{start})} 
\cdot \frac{8}{\widetilde H_f^2\mathrm{Re}}\right] \, .
\end{equation}

We performed tests for two parameter sets (Table~\ref{tab:THparams}). Set I corresponds
to the FSI3 benchmark in \cite{turek2006proposal}. It has matching fluid and solid densities and we employ
a Saint-Venant--Kirchhoff material model. Set II uses different material properties for the circle and the rectangular tail.
This last numerical example demonstrates that our FSI framework can also handle non homogenous material properties for the solid structure.

\begin{table}[htbp]
\centering
\caption{Parameters of Turek--Hron runs}
\begin{tabular}{|l|c|c|}
\hline
parameters & I (FSI3) & II  \\
\hline
$\rho_s$ $[\si{\kilogram\per\cubic\meter}]$ & 1000 & 1000  \\
\hline
$\mu_s$ $[\si{\mega\pascal}]$ & 0.5 & circle: 2, tail: 0.1  \\
\hline
$\lambda_s$ $[\si{\mega\pascal}]$ & 4.67 & circle: 4.67, tail: 0.23 \\
\hline
$\rho_f$ $[\si{\kilogram\per\cubic\meter}]$ & 1000 & 1000 \\
\hline
$\mu_f$ $[\si{\pascal\cdot\second}]$ & 1 & 1 \\
\hline
$U_{\mathrm{ref}}$ $[\si{\meter\per\second}]$  & 2 & 2  \\
\hline
\end{tabular}
\label{tab:THparams}
\end{table}

\begin{figure}[htbp]
\begin{center}
\includegraphics[width=\textwidth]{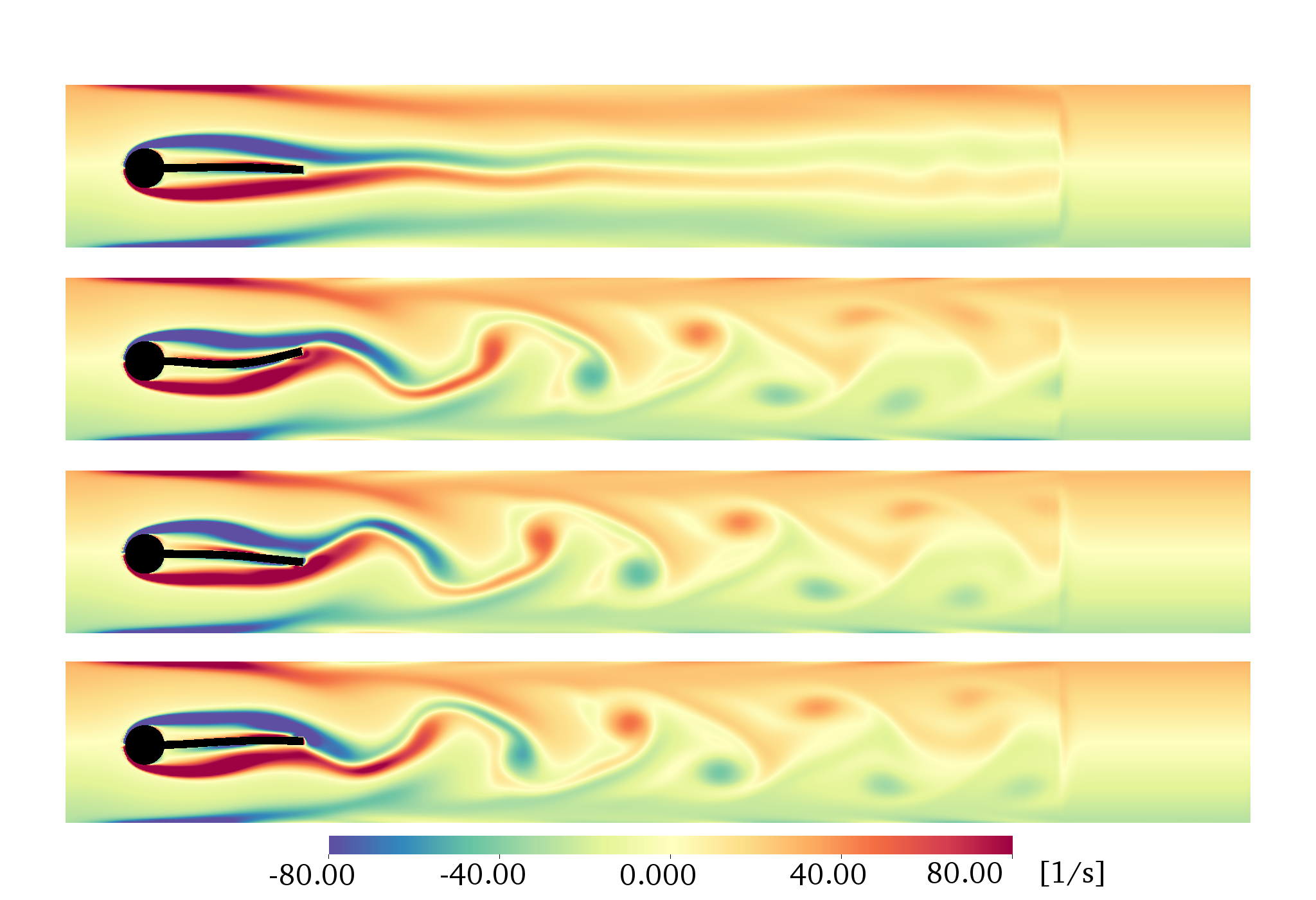}
\end{center}
\caption{Fluid vorticity $[\si{\per\second}]$ at times $t=1.34,\,2.77,\,5.50,\,19.17\,
[\si{\second}]$ for parameter set I. Times correspond to snapshots at $7,\,14.5,\,28.75$ and 
$100.25$ steady-state periods.}
\label{Vorticity_TH}
\end{figure}

Figure~\ref{Vorticity_TH} illustrates the temporal evolution of the vorticity field at different points in time.
The effect of the fringe forcing can be observed at the downstream end of the computational domain: the vortices are damped out and the 
Poiseuille flow is re-established.

\begin{figure}[htbp]
\centering
\begin{subfigure}[t]{0.49\textwidth}
\centering
\includegraphics[width=\textwidth]{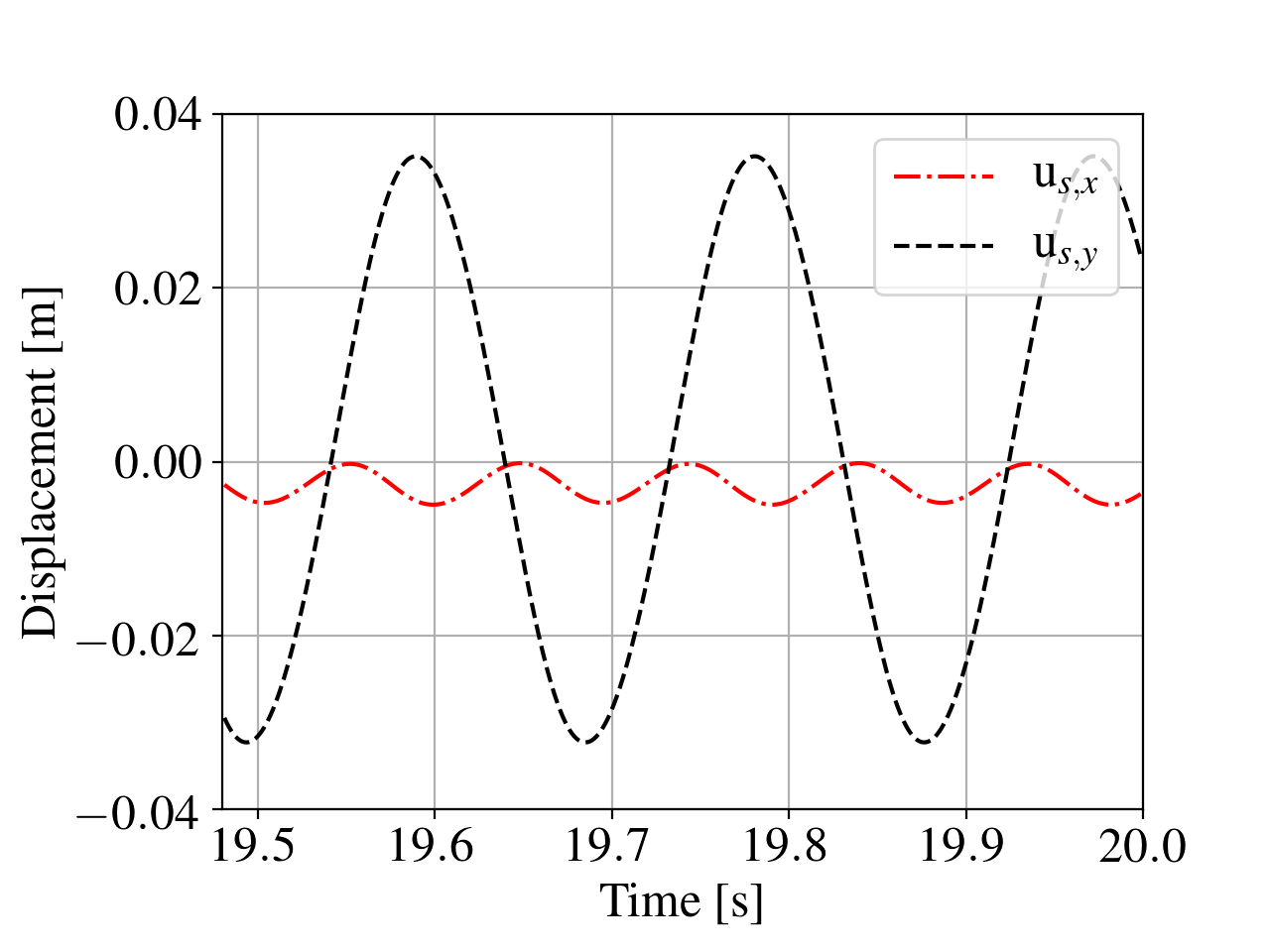}
\caption{}
\label{fig:FSI3disp}
\end{subfigure}
\begin{subfigure}[t]{0.49\textwidth}
\centering
\includegraphics[width=\textwidth]{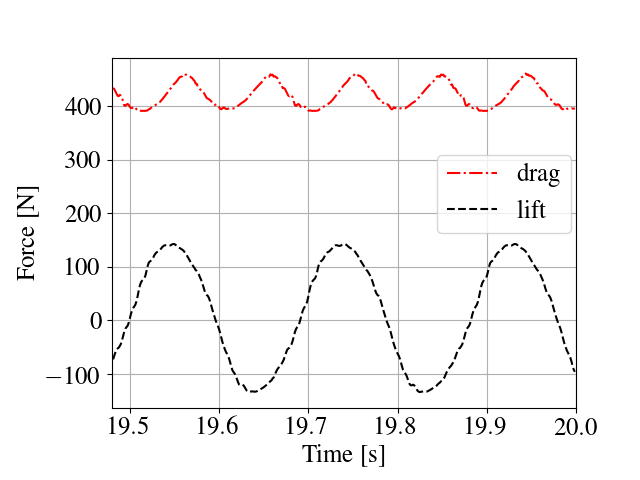}
\caption{}
\label{fig:FSI3liftdrag}
\end{subfigure}
\caption{(\subref{fig:FSI3disp}) Displacements of control 
point A located at the end of the elastic beam and (\subref{fig:FSI3liftdrag}) lift 
and drag forces for parameter set I (Table~\ref{tab:THparams}).}
\label{fig:THdisp}
\end{figure}

Figure~\ref{fig:FSI3disp} shows the displacements in $x$- and  
$y$-direction of a control point 
$A = (0.6\,[\si{\meter}], 0.2\,[\si{\meter}])$ located at the end of 
the elastic tail (Figure~\ref{fig_bench_TH}). 
The quantities of interest were fitted with a sinusoidal function of the form
$[A\cdot\sin(2\pi\cdot f\cdot t 
+ \phi) + M]$ and the retrieved values can be found in Table~\ref{tab:THstats}. The mean vertical displacement $M$ is 
$0.00137$ with amplitude $A=\pm 0.0338\,[\si{\meter}]$ and the horizontal displacement is $-0.00255\pm 
0.00231\,[\si{\meter}]$; 
the frequency $f$ of the $y$-displacement $u_{s,y}$ is $5.23\, [$\si{\hertz}$]$, 
and the frequency $f$ for the $x$-displacement $u_{s,x}$ is 
about $10.45\,[$\si{\hertz}$]$. 
The drag and lift forces on the structure over time are presented in Figure~\ref{fig:FSI3liftdrag} and their values can be found in Table~\ref{tab:THstats}.
All quantities agree well with the results obtained with other numerical 
methods applied to the same problem~\cite{turek2006proposal}.

\begin{figure}[htbp]
\centering
\begin{subfigure}[t]{0.49\textwidth}
\centering
\includegraphics[width=\textwidth]{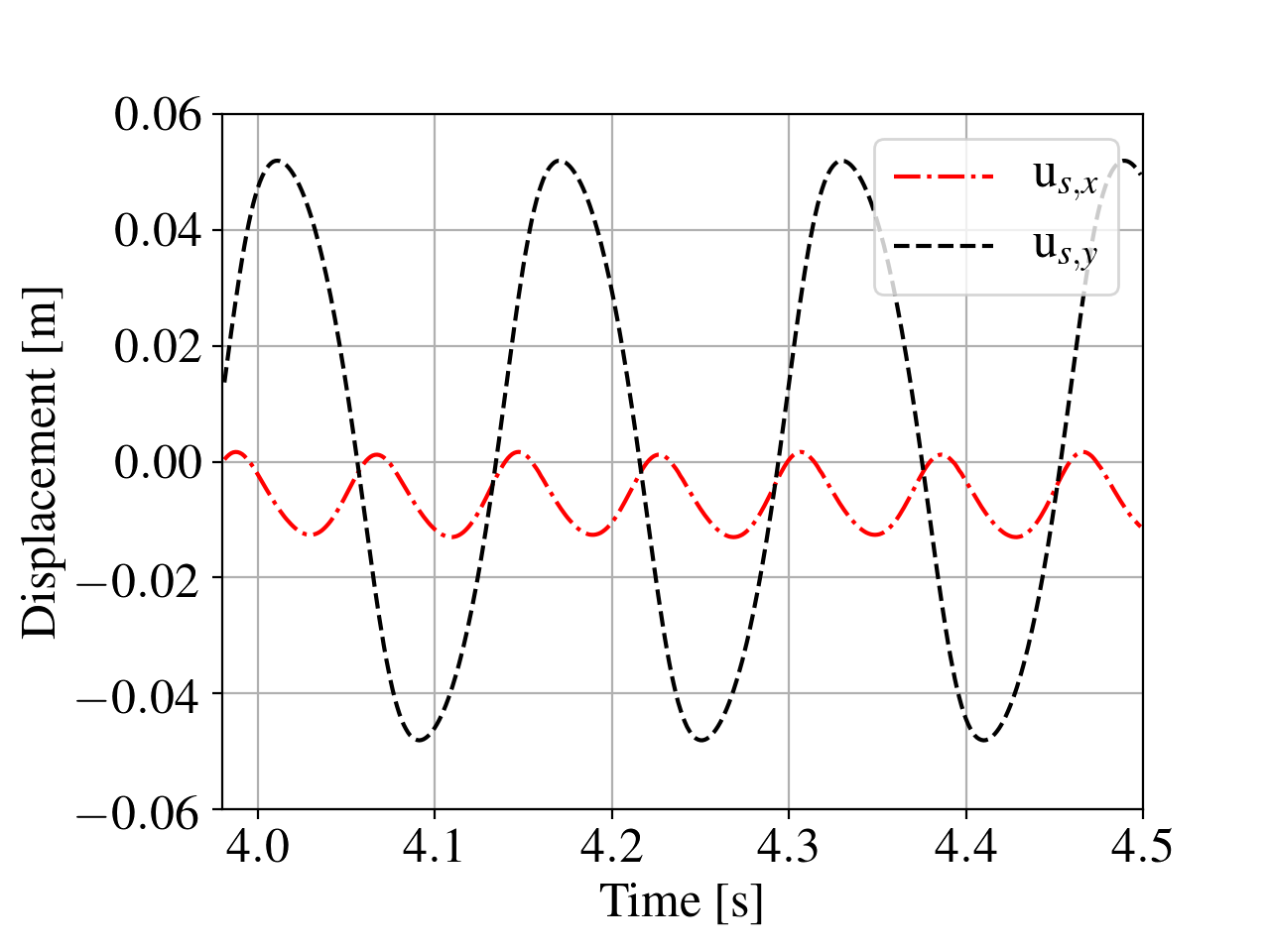}
\caption{}
\label{fig:softdisp}
\end{subfigure}
\begin{subfigure}[t]{0.49\textwidth}
\centering
\includegraphics[width=\textwidth]{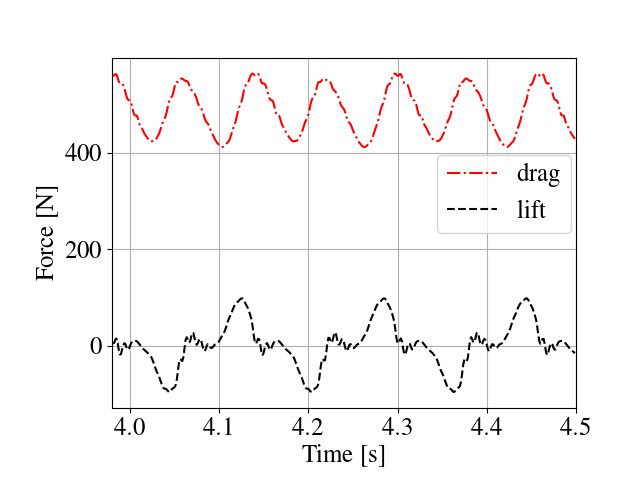}
\caption{}
\label{fig:softforce}
\end{subfigure}
\caption{(\subref{fig:softdisp}) Displacements of control 
point A at the end of the elastic beam and (\subref{fig:softforce}) lift 
and drag forces for 
parameter set II (Table~\ref{tab:THparams}).}
\label{fig:THsoft}
\end{figure}

We further consider another set of parameters (set II, Table~\ref{tab:THparams}) with inhomogeneous mechanical 
properties for the circle (C) and the rectangular tail (T) of the solid beam. 
The results in Figure~\ref{fig:softdisp} show an amplitude $A$ of $0.0513\,
[\si{\meter}]$ for the 
vertical displacement and of $0.00707\,[\si{\meter}]$ for the mean 
horizontal displacement. Moreover, the frequency $f$ is 
$6.27\,[$\si{\hertz}$]$ for the $y$-displacement $u_{s,y}$ and 
$12.54\,[$\si{\hertz}$]$ for the $x$-displacement $u_{s,x}$.  Lift and drag forces are visualized in Figure~\ref{fig:softforce}, and the corresponding quantities of interest, i.e. amplitude and frequency, can be found in Table~\ref{tab:THstats}.

\begin{table}[htbp]
\caption{Values of oscillating quantities obtained by a fit of the form 
$[A \cdot\sin(2\pi\cdot f \cdot t 
+ \phi) + M]$ }
\centering
\begin{tabularx}{\textwidth}{| l | X | X | X | X | X | X |}
\hline
Sets & \multicolumn{3}{c|}{I (FSI3)} & \multicolumn{3}{c|}{II}  \\
\hline
& $M$ $[\si{\meter}]$ & $A$ $[\si{\meter}]$ & $f$ $[\si{\hertz}]$ 
& $M$ $[\si{\meter}]$ & $A$ $[\si{\meter}]$ & $f$ $[\si{\hertz}]$\\
\hline
$u_{s,x}$  & 
$\num{-2.55e-3}	$ & 
$\num{2.31e-3}$ & 
$10.45 $ &
$\num{-6.21e-3}	$& 
$\num{7.07e-3}$& 
$12.54 $\\
\hline
$u_{s,y}$& 
$\num{1.37e-3}$& 
$\num{3.38e-2}$& 
$5.23$ &
$\num{1.83e-3}$& 
$\num{5.13e-2}$& 
$6.27$\\
\hline
& $M$ $[\si{\newton}]$ & $A$ $[\si{\newton}]$ & $f$ $[\si{\hertz}]$
& $M$ $[\si{\newton}]$ & $A$ $[\si{\newton}]$ & $f$ $[\si{\hertz}]$\\
\hline
drag &
$\num{421.10}$ & 
$\num{31.28}$ & 
$10.44$&
$\num{489.02}$ &
$\num{67.63}$ &
$12.54$ \\
\hline
lift & 
$\num{3.95}$ & 
$\num{140.68}$& 
$5.23$&
$\num{1.755}$ &
$\num{56.17}$ &
$6.24$\\
\hline
\end{tabularx}
\label{tab:THstats}
\end{table}

\subsubsection{Convergence Studies}
The Turek--Hron FSI3 benchmark is solved on a series of refined meshes to study convergence in space. 
The fluid domain is discretized on an $M \times N$ cartesian grid, and the Lagrangian domain is discretized using a mesh of linear elements ($\mathbb P_1$) with a  space discretization step equal to $h_{si}=
\{1.5\cdot10^{-3},\,3.10\cdot10^{-3},\,5.2\cdot10^{-3}, 1.0\cdot10^{-2}\}[\si{\meter}]$. The sizes of the resulting fluid grids and solid meshes are reported in Table~\ref{TableConv}.

\begin{table}[htbp]
\centering
\caption{Fluid Grid and Solid Mesh refinements.}
\label{TableConv}
\begin{tabular}{|l|c|c|}
\cline{1-3}
                & Fluid Grid Points       &   Solid Mesh Elements/Nodes     \\
 \cline{1-3}
 (i=1)\,coarse      &   $\,\,576\times\,\, 96$              &   $\,\,\,\,479/\,\,\,\,310$   \\  
 \cline{1-3}
 (i=2)\,medium    &   $\,\,769\times128$                 &   $\,\,1179/\,\,\,\,686$   \\ 
 \cline{1-3}
 (i=3)\,fine           &   $1153\times193$                   &   $\,\,3286/\,\,1805$    \\ 
 \cline{1-3}
 (i=4)\,finest        &   $2304\times384$                   &   $14753/\,\,7710$  \\ 
 \cline{1-3}
  \cline{1-3}
 \,reference        &   $4608\times762$                   &   $44259/23130$  \\ 
 \cline{1-3}
\end{tabular}
\end{table}

\begin{figure}[htbp]
\begin{subfigure}[t]{0.48\textwidth}
\centering
\includegraphics[width=\textwidth]{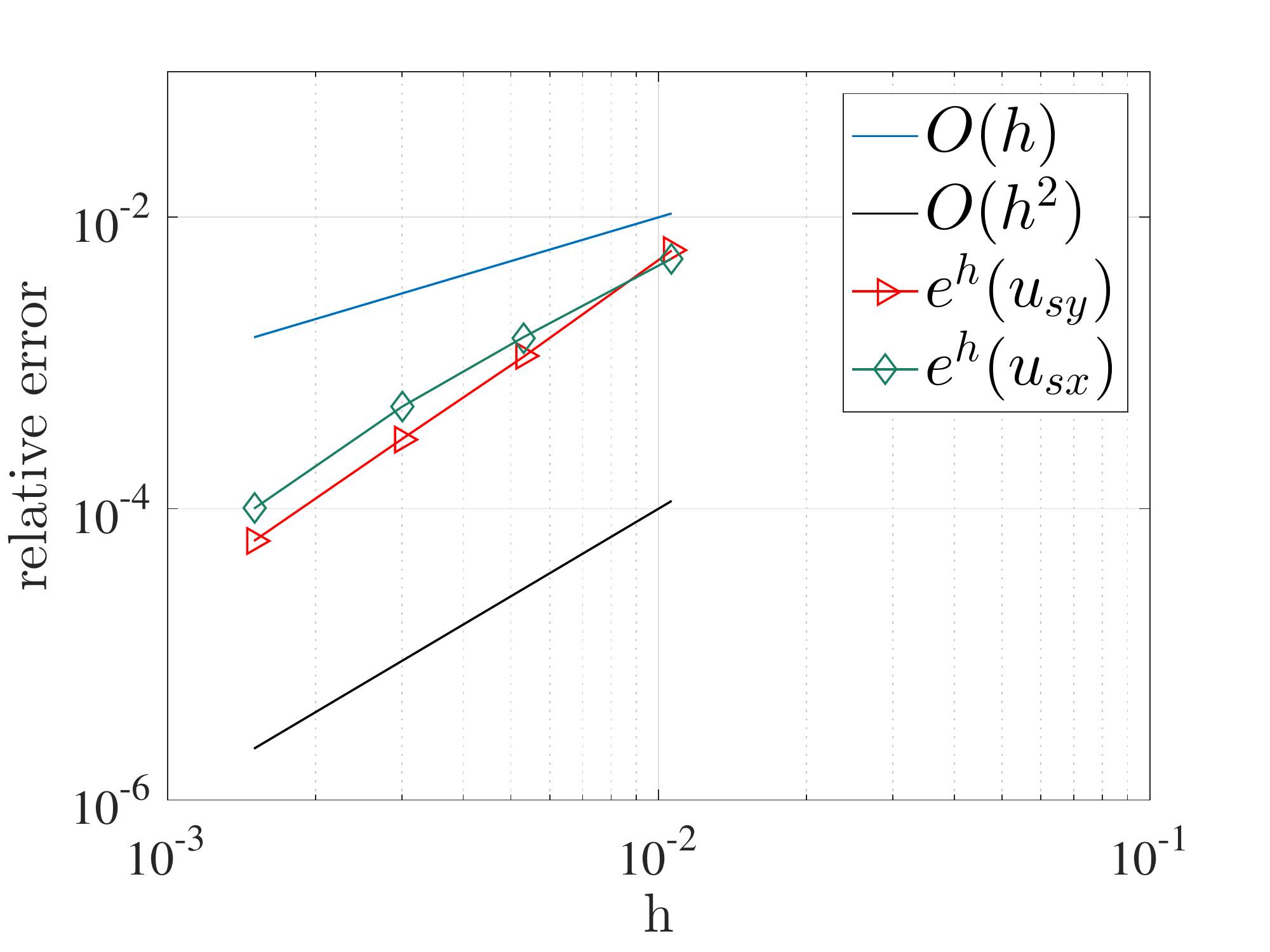}
\caption{Solid displacement errors.}
\end{subfigure}
\begin{subfigure}[t]{0.48\textwidth}
\centering
\includegraphics[width=\textwidth]{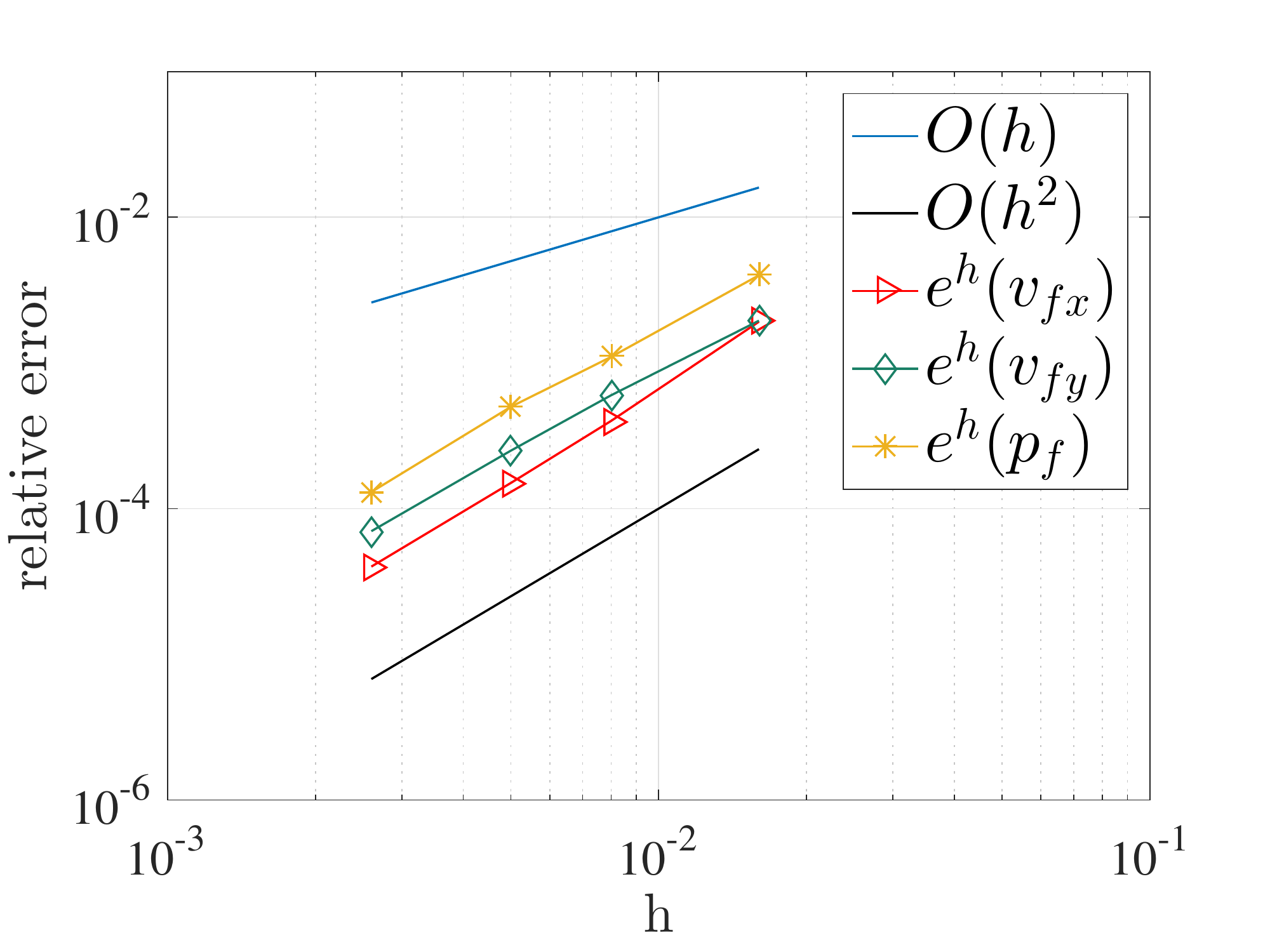}
\caption{Fluid velocity and pressure errors.}
\end{subfigure}
\caption{Relative errors $e^h(u_{sx})$, 
$e^h(u_{sy})$,
$e^h(v_{fx})$,
$e^h(v_{fx})$ and 
$e^h(p_{f})$ in space of $x$- and $y$-components of solid displacements $\mathbf{u}_s$ and 
fluid velocities $\mathbf{v}_f$ and pressure $p_f$, respectively. Lines 
with slopes of  $O(h)$ and $O(h^2)$ are shown for reference.}
\label{conv_space}
\end{figure}

A relative $L^2$-norm error $e^h(\theta^h)=||\theta^h-\theta^r||_2/||\theta^r||_2$ of a 
generic variable $\theta^h$ is computed with respect 
to the reference solution $\theta^r$ obtained with the highest resolution (Table~\ref{TableConv}). 
As can be observed in Figure~\ref{conv_space} the 
displacement field shows a convergence rate between first and second order, whereas fluid pressure and velocity fields converge nearly quadratically.

\begin{figure}[htbp]
\begin{subfigure}[t]{0.48\textwidth}
\centering
\includegraphics[width=\textwidth]{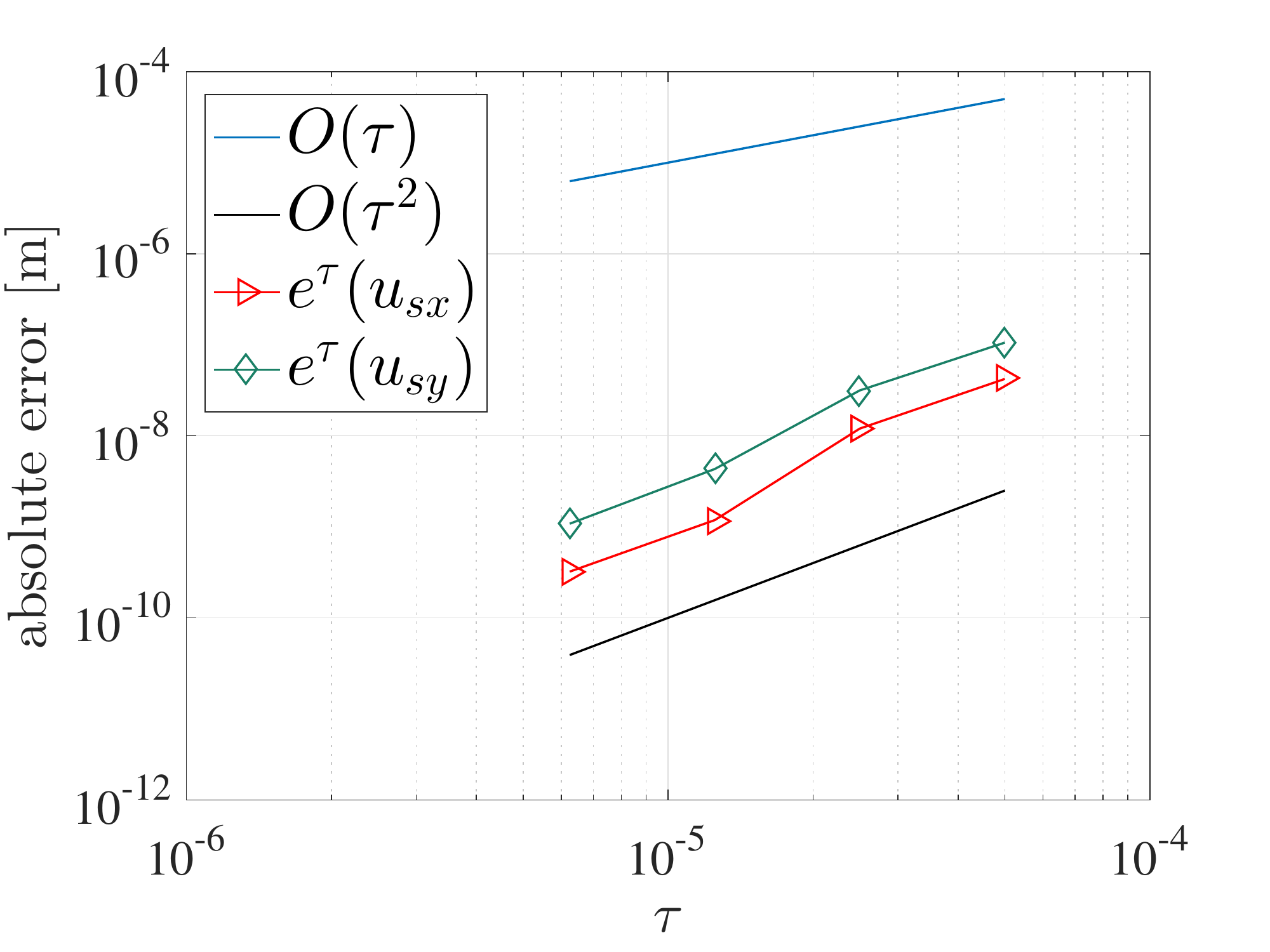}
\caption{Solid displacement errors.}
\end{subfigure}
\begin{subfigure}[t]{0.48\textwidth}
\centering
\includegraphics[width=\textwidth]{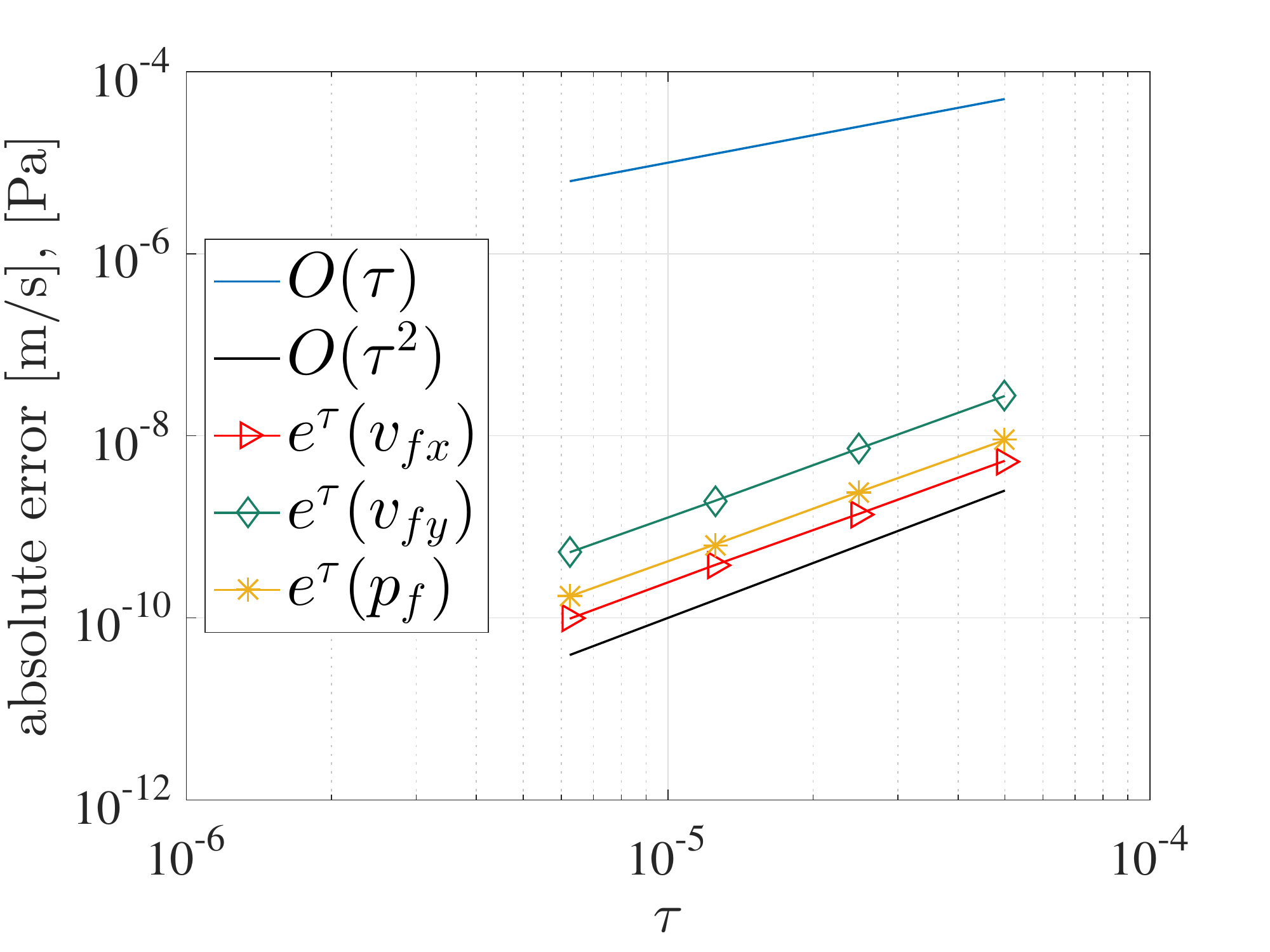}
\caption{Fluid velocity and pressure errors.} 
\label{sfig: fluidabs}
\end{subfigure}
\caption{Absolute errors $e^\tau(u_{sx})$,
$e^\tau(u_{sy})$,
$e^\tau(v_{fx})$,
$e^\tau(v_{fy})$ and
$e^\tau(p_{f})$
in time of $x$- and $y$-components of solid displacements $\mathbf{u}_s$ and 
fluid velocities $\mathbf{v}_f$ and pressure $p_f$ in the  $L^2$-norm 
for $t=4.8\,[\si{\second}]$. Lines with slopes of $O(\tau)$ and 
$O(\tau^2)$ are shown for reference.}
\label{conv_time}
\end{figure}

The convergence in time is analyzed by adopting the finest spatial grid for fluid and solid and a 
time discretization parameter $\Delta t$ equal to  $\{5\cdot10^{-5},\,2.5\cdot10^{-5},\, 1.25\cdot10^{-5},\, 0.625\cdot10^{-5}\}\,
[\si{\s}]$. An $L^2$-norm error $e^\tau(\theta^\tau) = {||\theta^{\tau}-\theta^r||_2}$ is computed at 
$t = 4.8\,[s]$ for a generic variable $\theta^\tau$ with respect to a reference solution 
$\theta^r$, obtained by using a time step of $0.15 \cdot10^{-5}\,[\si{\second}]$.
The results in Figure~\ref{conv_time} show that second order convergence rate is obtained for the variables.

\subsection{Flow-induced oscillation of an inert plate}
\label{ssec:oscillation}

In this section, we present a benchmark for the treatment of inertial forces. 
This benchmark has an exact analytical solution (details are given in the \ref{sec:inertiabench}). 
It consists of an infinitely long plate of thickness $2\delta_s$ immersed in a Newtonian incompressible 
fluid driven by an oscillating pressure gradient.
The wall shear stresses on the surface of the plate induce an oscillation of the plate. For very 
light plates and slow 
frequencies the plate will oscillate synchronously with the fluid. For very heavy plates and high 
frequencies, the plate 
will experience only small oscillation amplitudes and a significant phase lag. It can be shown that 
the oscillation of the plate (amplitude and 
phase lag) is governed by a single parameter $\beta$ (Equation~\eqref{eq:beta}) which is built with the solid/fluid 
density ratio $\rho_s/\rho_f$ and the Womersley number $\delta_s \sqrt{(2\pi f_0/\nu_f)}$, 
where $\nu_f=\mu_f/\rho_f$ is the kinematic viscosity. 

\begin{figure}[htb]
\centering
\def\svgwidth{\textwidth}
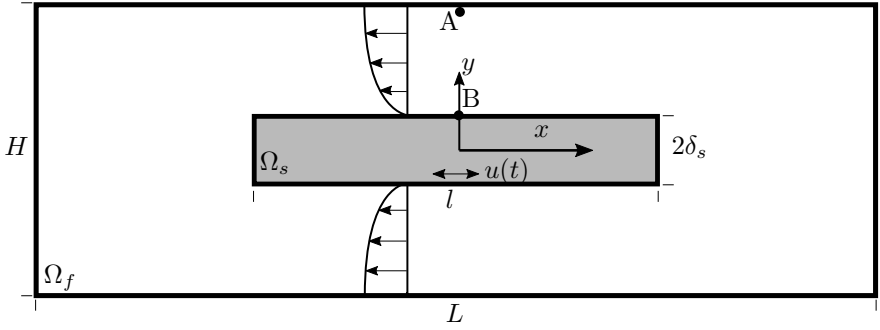
\caption{\label{fig15}Geometry for the flow-induced oscillation benchmark. Data is sampled at 
extraction points A (fluid) and B (solid).}
\end{figure}

In the numerical experiments presented herein, the infinitely long plate is modelled as a 
two-dimensional beam with finite length
$l\gg \delta_s$ located in the middle of the fluid channel as shown in Figure~\ref{fig15}. 
We set $l=1\,[\si{\meter}]$ and  
$\delta_s=0.006\,[\si{\meter}]$, whereas the dimensions of the fluid domain are: 
length $L=2\,[\si{\meter}]$, height $H = 0.2\,[\si{\meter}]$. The center of the 
plate and of the fluid domain is located at $y=0$.
Periodic boundary conditions are imposed on all boundaries of the fluid domain, whereas a forcing 
term $\partial p /\partial x= G \cos(2\pi f_0 t)$ is applied over the entire domain with frequency $f_0=10 [\si{\hertz}]$ 
and $G=110500[\si{\pascal}]/2.2[\si{\meter}] = 50227[\si{\pascal\per\meter}]$. To this aim, 
the nondimensional forcing term in the fringe region is modified as follows:
\begin{equation}
\widetilde{\bm{f}}_\mathrm{fringe} = 
\left[G\cdot \frac{L_\mathrm{ref}}{\rho_f U_\mathrm{ref}^2} 
\right]\cos(2\pi f_0 L_\mathrm{ref}/U_\mathrm{ref} \widetilde t)\, .
\end{equation}

In order to study the motion of the inert plate as a function of the parameter $\beta$, we consider physical 
parameters for three different test cases as indicated in Table~\ref{tab:TestCases} and model the solid material as linear 
elastic with  $E = 20.0\, [\si{\mega\pascal}]$  and $\nu=0.4$. Only reaction forces on the top and 
bottom edges of the plate are communicated in order to exclude normal forces on the edges that 
are normal to the flow. These edges are not present in the analytic formulation of the problem
where an infinitely long domain is assumed.

\begin{table}[htb]
\centering
\caption{Parameter settings for the flow-induced oscillations of a inert plate benchmark.}
\label{tab:TestCases}
\begin{tabular}{|l|l|l|l|l|}
\cline{1-5}
$\text{test case}$& $\rho_s\,[\si{\kilogram\per\cubic\meter}]$ & $\rho_f\,[\si{\kilogram\per\cubic\meter}]$ & $\mu_f\,[\si{\kilogram\per\meter\square\second}]$   &  $\beta$       \\ 
\cline{1-5}
        A                 &   \,\,100    &   1000    &   1.0         &   \,\,0.106     \\  
\cline{1-5}
        B                 &   1000      &   1000    &   1.0        &    \,\,1.06       \\ 
\cline{1-5}
        C                 &   1000      &  1000     &   0.01      &    10.6           \\ 
 \cline{1-5}
\end{tabular}
\end{table}

\begin{figure}[htbp]
\centering
\begin{subfigure}[t]{\textwidth}
\centering
\includegraphics[width=0.49\textwidth]{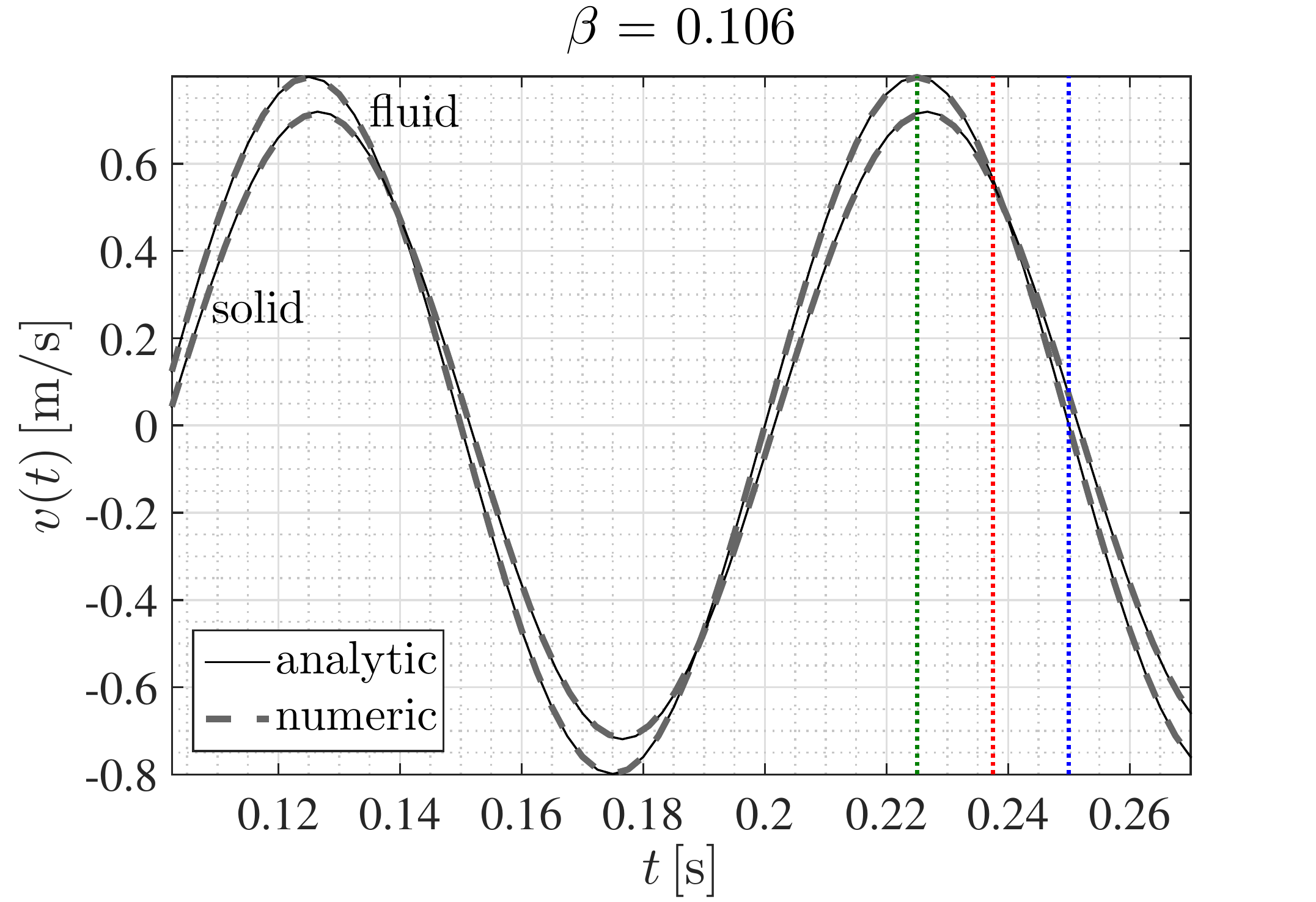}
\includegraphics[width=0.49\textwidth]{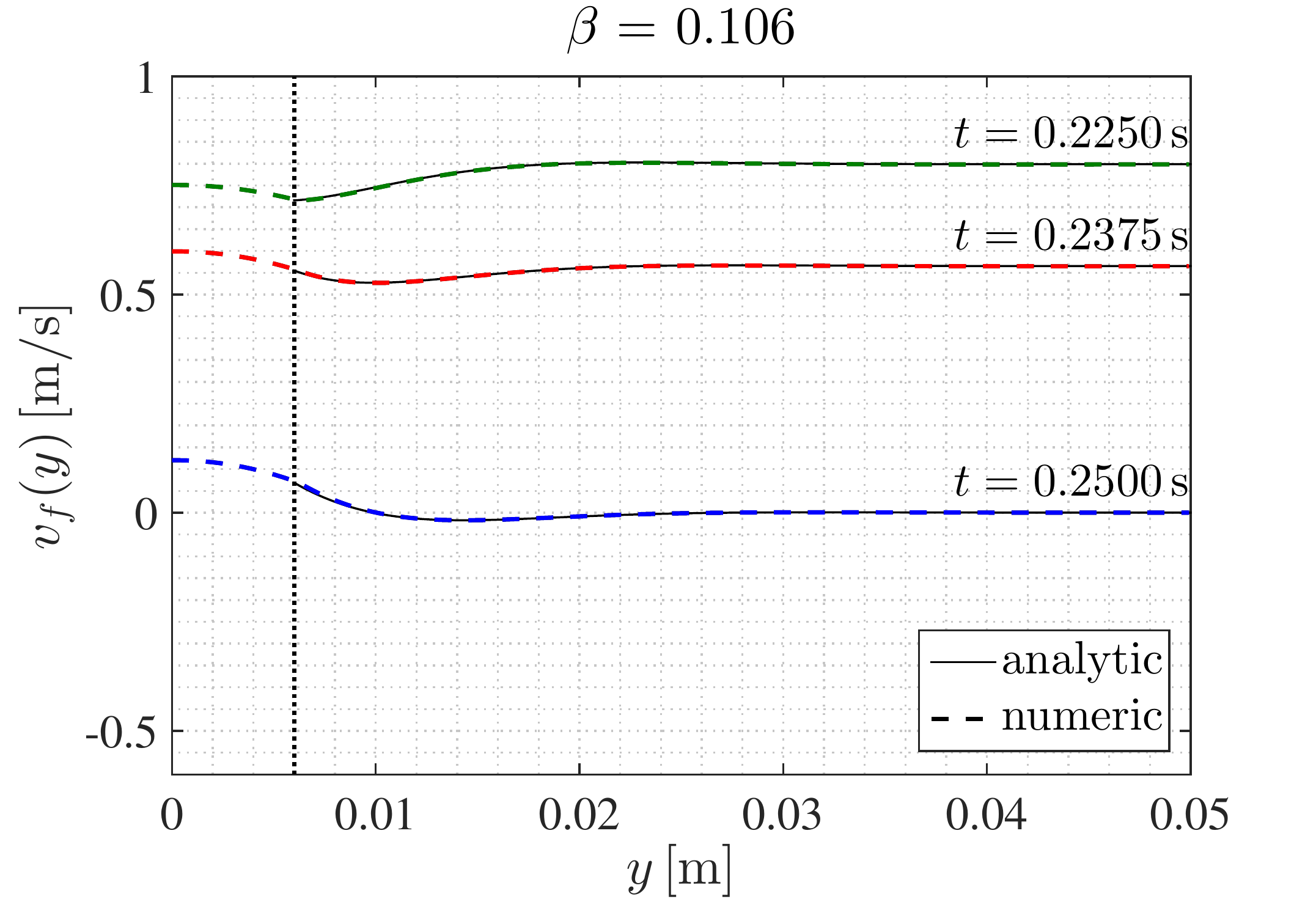}
\caption{Test case A}
\label{fig:beta01a}
\end{subfigure} 
\begin{subfigure}[t]{\textwidth}
\centering
\includegraphics[width=0.49\textwidth]{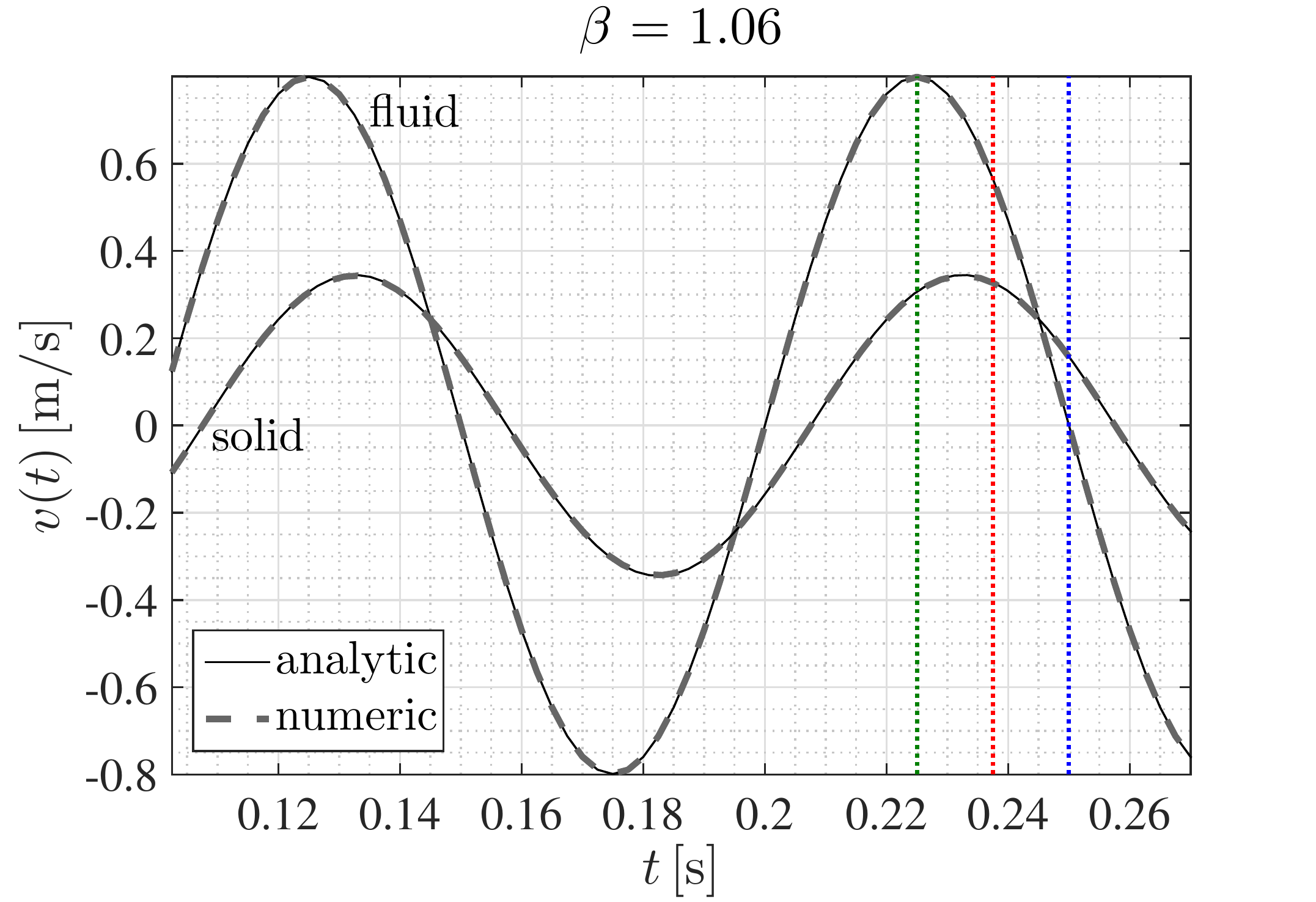}
\includegraphics[width=0.49\textwidth]{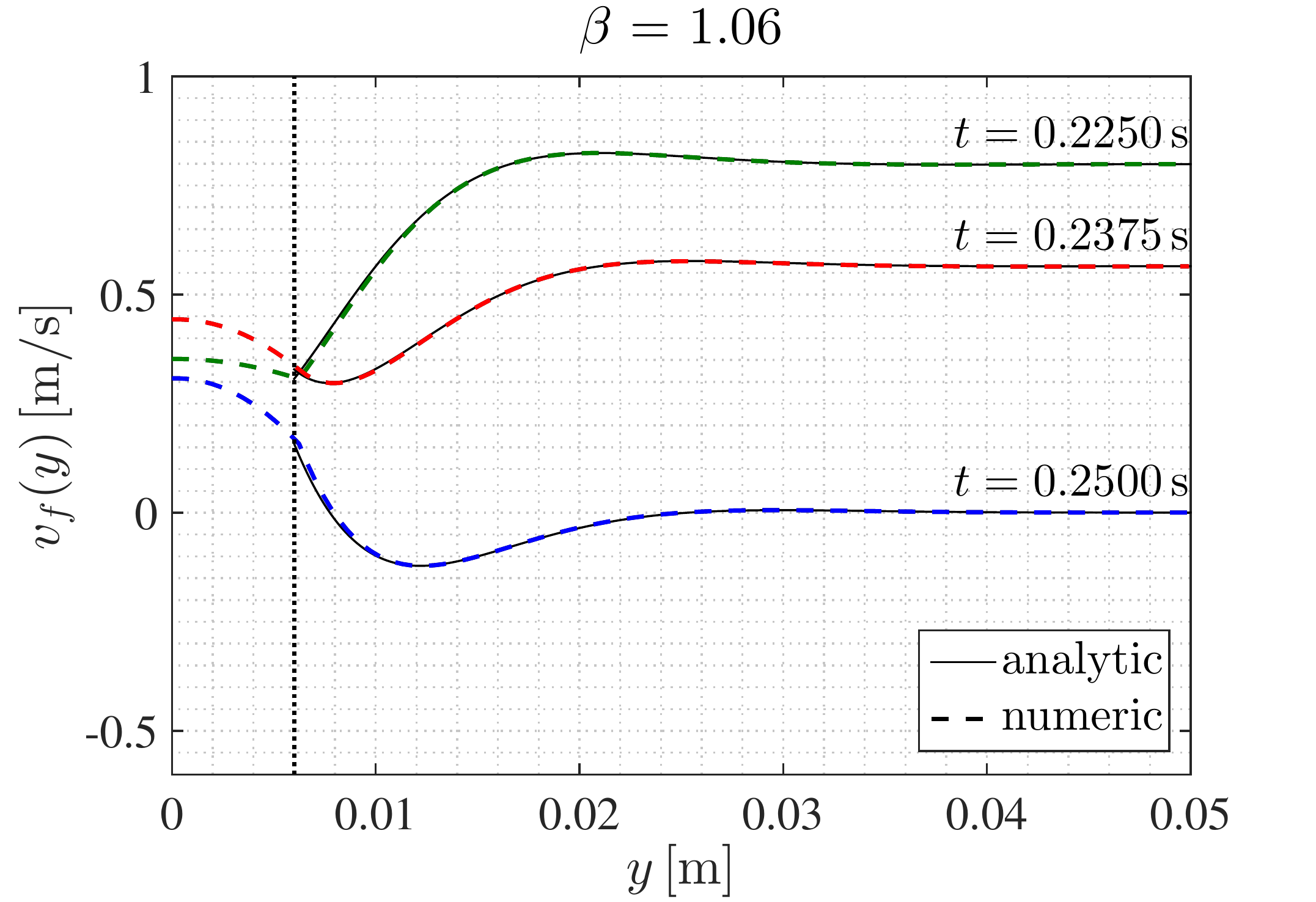}
\caption{Test case B}
\label{fig:beta1a}
\end{subfigure}
\begin{subfigure}[t]{\textwidth}
\centering
\includegraphics[width=0.49\textwidth]{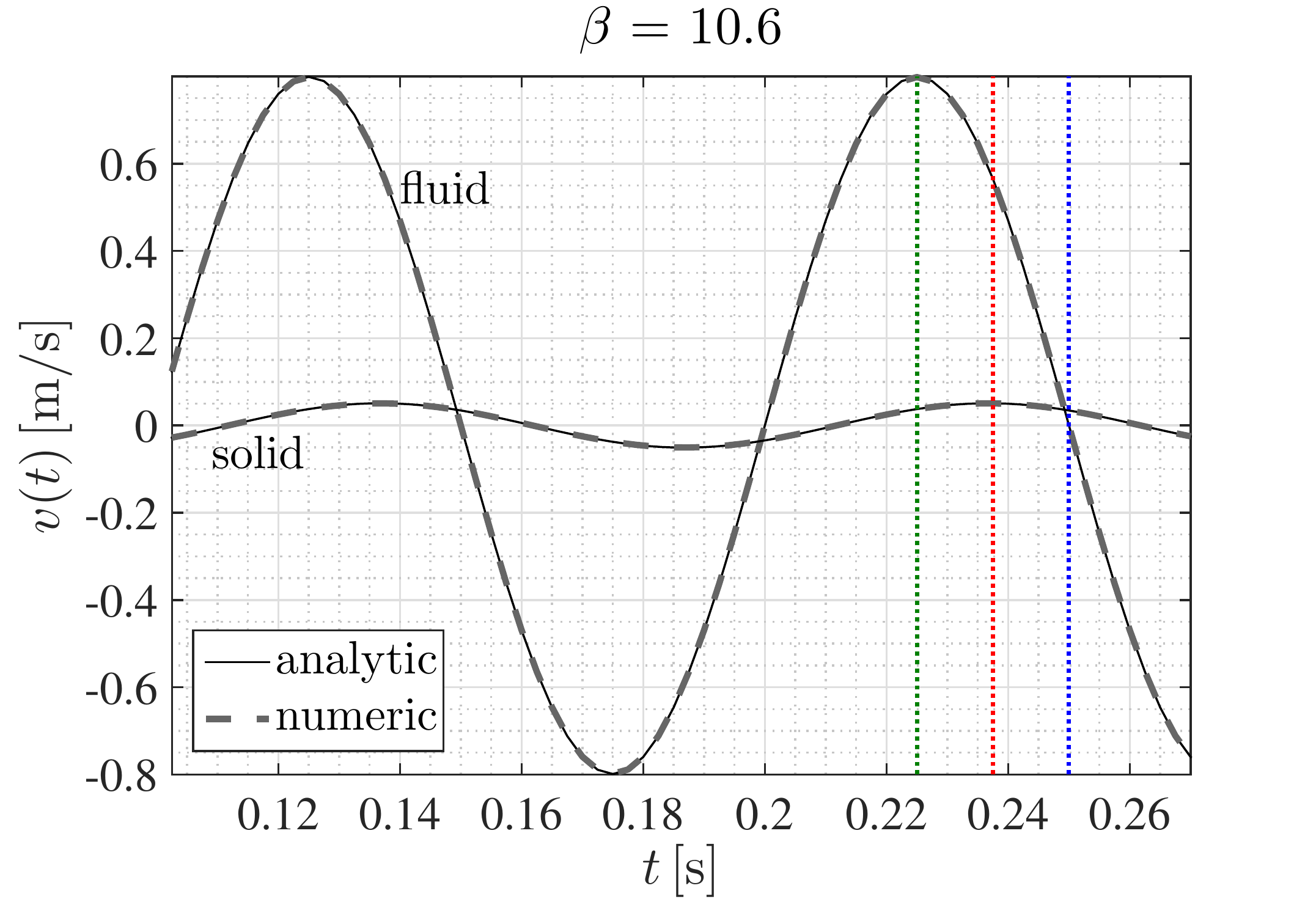}
\includegraphics[width=0.49\textwidth]{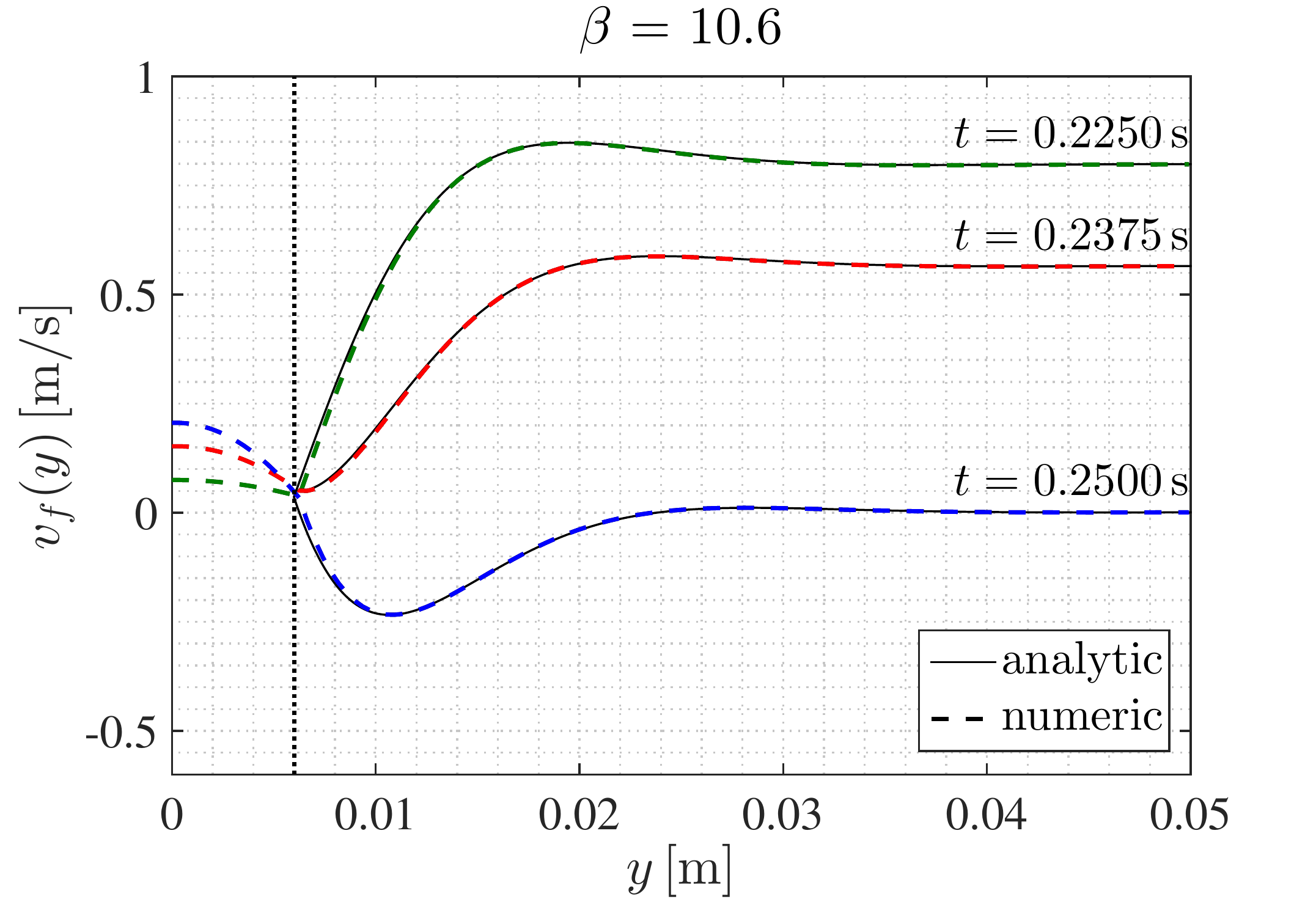}
\caption{Test case C}
\label{fig:beta10a}
\end{subfigure}
\caption{Velocity profiles over time and space of the test cases summarized in
Table~\ref{tab:TestCases}.
Left: Fluid and solid velocities over time at probe points A and B (Figure~\ref{fig15}), 
respectively.  Right: Fluid 
velocity profiles in the middle of the flow channel. Solid 
lines refer to the analytical solution and dashed lines refer to the numerical values. The time instances at which the velocity profiles were 
extracted are highlighted in their respective color in the plots on the left. Numerical results for the fluid velocity are also plotted within
the solid domain $y<\delta_s$. }
\label{fig:inertia_bench}
\end{figure}

We use $\mathbb{P}_1$ finite elements for the space discretization of the solid subproblem and,
as for the Turek--Hron benchmark, attach bilinear basis functions ($\mathbb{Q}_1$) to the fluid grid.
The fluid domain is discretized using
a $2048\times129$ Cartesian grid whereas the solid domain is discretized with a triangular mesh with a space discretization step equal to $h_s=0.001$.

Figure~\ref{fig:inertia_bench} (left) shows the fluid velocity $v_{f}$ at point A and solid 
velocities at point B for all three test cases ($\beta=0.106,
\,1.06,\,10.6$).
The numerical results show a good agreement with the 
analytical solution (solid lines) for fluid and structure.

On the right, Figure~\ref{fig:inertia_bench} shows the velocity profiles $v_f(y)$ 
obtained along a vertical section in the middle of the flow channel at times $t=0.225,\,0.2375,\,0.25\,[\si{\second}]$. Given the 
symmetry of the problem, we only show profiles for $y>0$. The black dotted line represents the location of the 
fluid-structure interface at $y=\delta_s$. 
One may observe that (I) a Stokes boundary layer is established near the inert 
plate, and that (II) each numerical velocity profile (dashed line) recovers the related analytical one 
(solid line) very well for all the three tests cases.

\begin{table}[htbp]
\centering
\caption{Comparison between the numerical and analytical values of the magnitude of the amplitude ratio $|\mathcal{A}|$ and of its phase-lag
$\phi{(\mathcal{A})}$ for the three test cases (Table~\ref{tab:TestCases}).}
\label{tab:TabComparison}
\begin{tabular}{|c|c|c|c|c|c|c|}
\cline{1-7}
$\beta$     & $\phi{(\mathcal{A})}\,[^{o}]$  & $\phi({\mathcal{A}}_{an})\,[^{o}]$ & $\frac{|\phi{(\mathcal{A}_{an})}-
\phi{(\mathcal{A})}|}{|\phi{(\mathcal{A}_{an})}|}$   & $|\mathcal{A}|$ & $|\mathcal{A}|_{an}$ & $\frac{|
\mathcal{A}|_{an}-|\mathcal{A}|}{|\mathcal{A}|_{an}}$ \\ 
\cline{1-7}
    0.106   &       \,\,-5.45    &      \,\,-5.47        &        $0.36\%$    &      0.898         &        0.90                      &           $ 0.22\%$  \\ \cline{1-7}
    1.06     &      -27.13  &      -27.22              &        $0.33\%$     &      0.4305         &        0.4316                  &           $ 0.25\%$  \\ \cline{1-7}
   10.6      &      -42.21
      &      -42.42              &        $0.49\%$     &      0.0633         &        0.0636                  &           $ 0.47\%$ \\ \cline{1-7}
\end{tabular}
\end{table}

Table \ref{tab:TabComparison} shows a quantitative comparison between the numerical results and the analytical solution (Equation~$
\ref{fig09}$) for the magnitude of the amplitude ratio ($| \mathcal{A} |$) and the phase-lag ($\phi(\mathcal{A})$). 
The results indicate a relative error less than $0.5\%$. 

\subsection{Towards cardiovascular applications: Elastic beam in a 3D fluid channel}
It is well known that human soft tissue is highly deformable and characterized by nonlinear and 
anisotropic 
behavior which requires the use of suitable anisotropic hyperelastic fiber-reinforced constitutive 
models. To this aim, 
we illustrate the capabilities of our FSI framework with a three-dimensional application consisting of 
an elastic beam immersed in a channel flow.
We employ the constitutive law proposed by Holzapfel (Equation~\eqref{eq:Holzapfel}) for modelling the 
anisotropic elastic behavior of human arterial walls and choose parameters for the fluid, which 
yield a Reynolds number regime typical pf large blood vessels. 

\begin{figure}[htbp]
\centering
\includegraphics[width=0.90\textwidth]{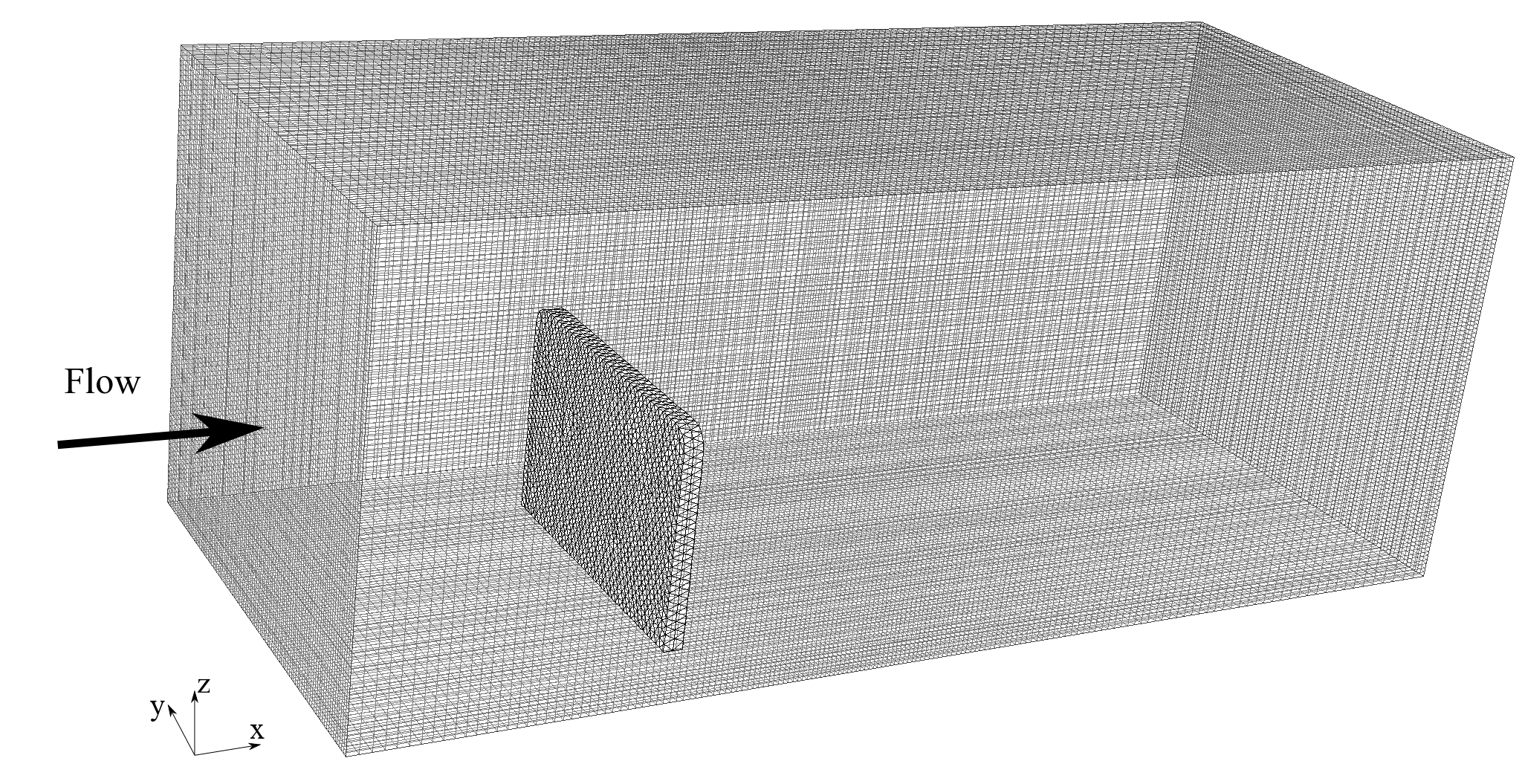}
\caption{Mesh Setup for the 3D benchmark. The fluid grid consists of 
$161\times97\times97$ $=1514849$ points, the solid mesh consists of $4555$ nodes and 
$19838$ tetrahedral elements. }
\label{fig:3D_100_fm}
\end{figure}

\begin{figure}[htbp]
\centering
\begin{subfigure}[t]{0.30\textwidth}
	\centering	
    \includegraphics[height=0.7\textheight]{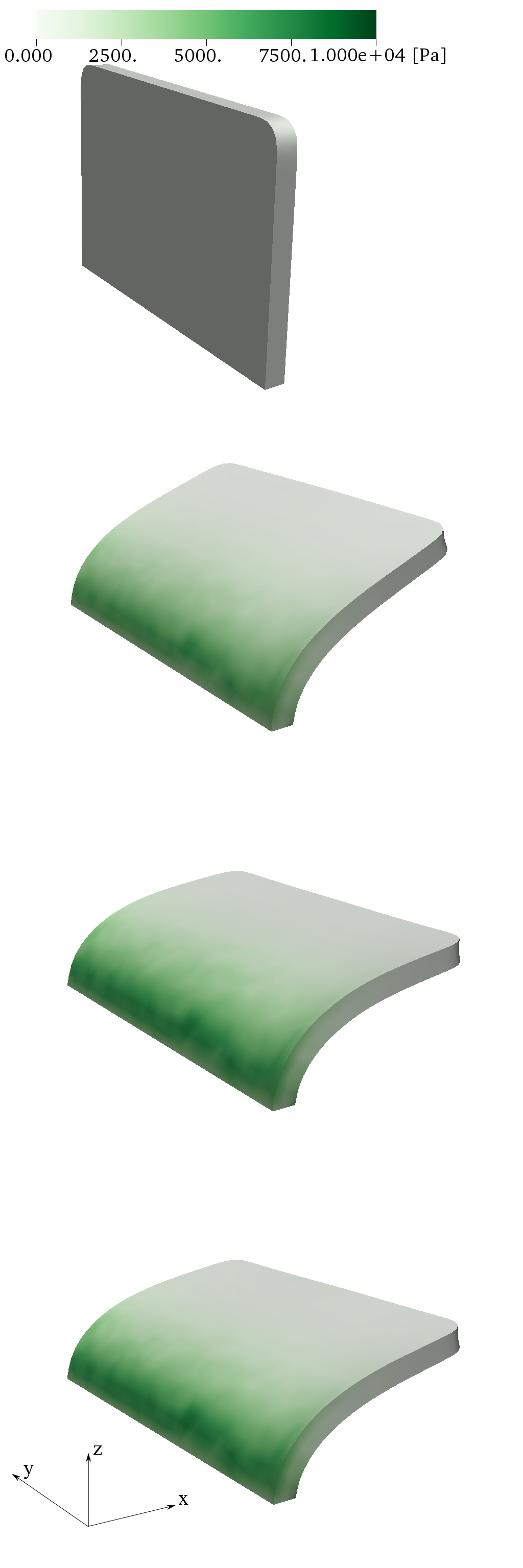}
	\caption{von Mises Stresses}
	\label{subf:vonMises}
\end{subfigure}
\begin{subfigure}[t]{0.68\textwidth}
	\centering
	\includegraphics[height=0.7\textheight]{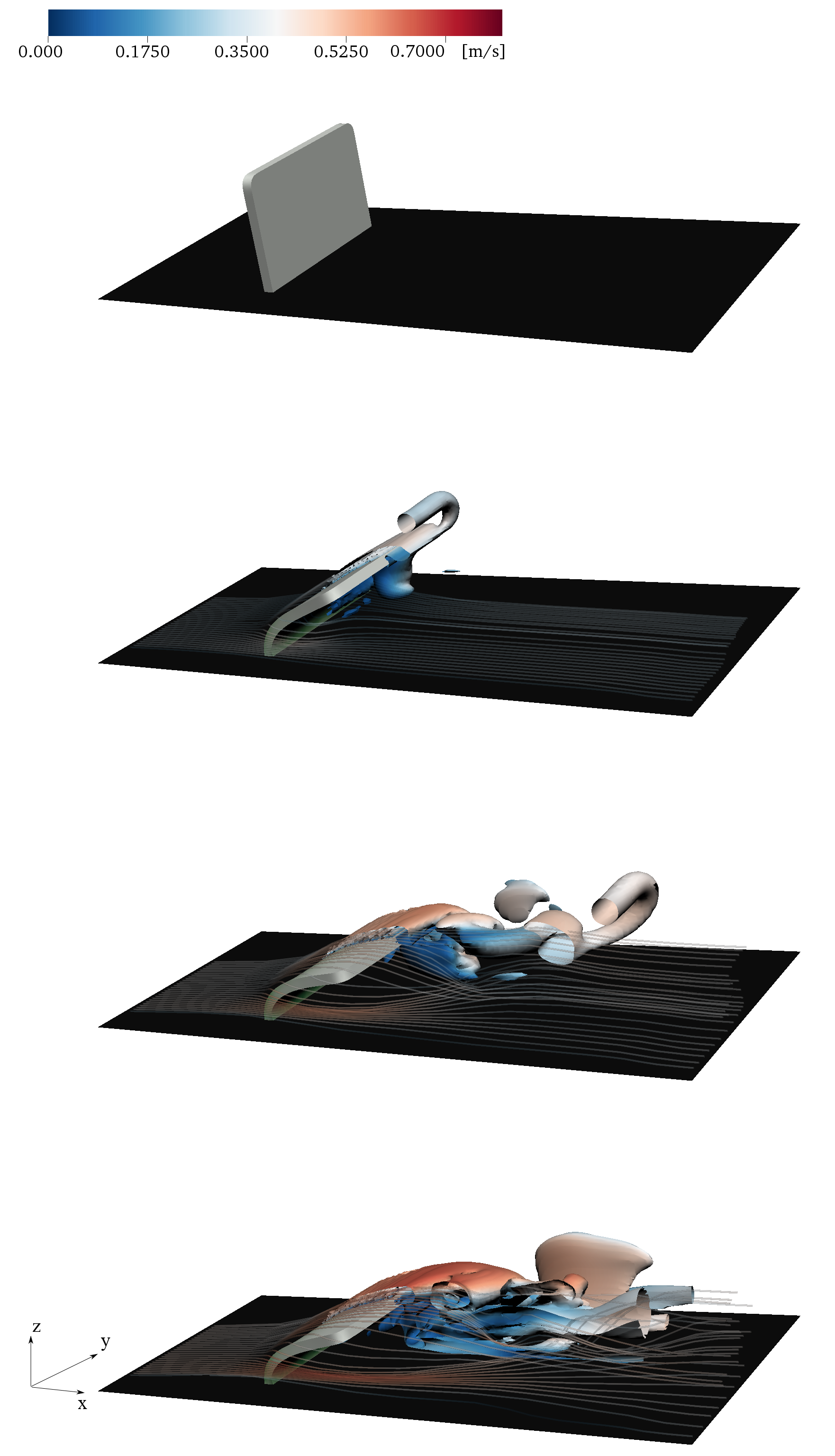}
	\caption{Streamlines and $\lambda_2$ isosurfaces}
	\label{subf:lambda2}
\end{subfigure}
\caption{Snapshots of the three-dimensional flexible membrane at times $t=0,0.055,0.11,0.18$\,$[\si{\second}]$ immersed in a channel flow at Re=2250. (\subref{subf:vonMises}) spatial 
distribution of the von Mises stresses of the solid structure. (\subref{subf:lambda2}) streamlines 
and isosurfaces for $\lambda_2=-0.54$ \cite{jeong1995identification} colored by the local velocity magnitude.}
\label{fig:beam_vel_mises}
\end{figure}

Let $\Omega_f$ be a three-dimensional fluid channel with extents $L_{f,x}=0.07\,[\si{\meter}]$, $L_{f,y}=0.03 \,[\si{\meter}]$ and 
$L_{f,z}=0.027 \,[\si{\meter}]$, and $\Omega_s$ be a parallelepiped with dimensions: $L_{s,x}=0.00115\,[\si{\meter}]$,  
$L_{s,y}=0.02\,[\si{\meter}]$ and  
$L_{s,z}=0.012\,[\si{\meter}]$ attached to the  bottom wall of the fluid channel at $x=0.02\,[\si{\meter}]$.

The fluid domain is discretized with a stretched Cartesian grid consisting of 
$161\times97\times97=1514849$ points, whereas the solid beam consists of $19838$ $
\mathbb{P}_1$ elements and $4555$ nodes (Figure~\ref{fig:3D_100_fm}). $
\mathbb{Q}_1$ basis functions are attached to the finite difference fluid grid for computing the transfer operator.
 
We set the fluid density to $\rho_f= 1000\,[\si{\kilogram\per\cubic\meter}]$, 
the reference length to $L_\mathrm{ref}=L_{f,z}=0.027[\si{\meter}]$ and the dynamic viscosity $
\mu_f=0.006\, [\si{\pascal\cdot\second}]$, which leads to a Reynolds number of $2250$.  
The solid structure and the fluid have the same density, and a nearly incompressible Holzapfel--Ogden material model is used
 with $\mu_s = 10\,[\si{\kilo\pascal}]$, $\kappa=1\,[\si{\mega\pascal}]
$, $k_{11}=10\,[\si{\kilo\pascal}]$ and $k_{12}=10\,
[\si{\kilo\pascal}]$. A local base ${\mathbf{e}_1,
\mathbf{e}_2,\mathbf{e}_3}$  is assumed with $\mathbf{e}_1$ oriented along the $z$-direction 
(i.e.\,$ \mathbf{g}_{01}=0\mathbf{e}_1+ 0\mathbf{e}_2+ 1\mathbf{e}_3$).

Periodic boundary conditions are imposed along the inlet and the outlet of the fluid channel 
together with no-slip boundary conditions on the lateral surfaces of the fluid 
channel. The flow is driven uniformly by a forcing region with parameters: 
$x_\mathrm{start}  = 0.0935\,[\si{\meter}]$, $x_\mathrm{end} = 
0.1\,[\si{\meter}]$ and $\hat \lambda=10$.

The mechanical response of the elastic beam is characterized by computing the von 
Mises stress ($\mathcal{VM}$)~\cite{leckie2009} which represent a scalar field quantity 
widely used to predict yielding failure of ductile materials subjected to any 
loading condition. 
The spatial distributions of $\mathcal{VM}$ at times $t=0,0.055,0.11,0.18$\,$[\si{\second}]$ depicted in Figure~\ref{subf:vonMises} 
show uniformly distributed stresses.
 
Finally, Figure~\ref{subf:lambda2} depicts the same beam together with streamlines and 
isosurfaces for vortical structures ($\lambda_2$-criterion, \cite{jeong1995identification}).
Both streamlines and isosurfaces are colored by the local velocity magnitude of the flow.
The flow evolves from rest to a transitional character with an initial starting vortex shed from
the top of the beam (see supplementary video). The starting vortex is advected out of the 
domain at which point more complex vortical structures start shedding from the top and the sides of 
the beam. 

\section{Conclusion}
\label{sec:conclusion}

In this paper we have proposed a novel FSI framework based on the IBM. The main contribution consists of three key advancements with respect to existing immersed methodologies.

First, the use of $L^2$-projections for the transfer of discrete data fields between nonconforming overlapping meshes ensures a modular 
and flexible coupling of independent flow and structure solvers based on different schemes for the time and space discretization. 
The current implementation of the variational transfer is based on piecewise affine meshes which allow for an efficient generation of the quadrature points for integrating in the intersection 
region. The proposed methodology can be extended to nonaffine meshes if high-order 
elements are adopted for the discretization of the solid subproblem.

Second, the description of the solid motion by the elastodynamics equations is solved via a 
fully implicit time integration scheme, which yields a robust  scheme for structural dynamics.

Third, the use of high-order finite difference methods for the flow solver 
allows for the DNS of laminar, transitional and turbulent flows interacting with complex structures. 

The benchmark results validate the framework with respect to the elastodynamics formulation, the flow solver, the fluid-structure 
interaction and the treatment of solid inertia. Furthermore, they demonstrate its capability of combining 
complex materials with flows at moderately high Reynolds numbers. The method is shown 
to be second-order accurate by means of a mesh- and time step-refinement study.

The present framework is thus able to solve the FSI of hyperelastic, nonhomogeneous and 
anisotropic structures, immersed in incompressible laminar, transitional or 
turbulent flow. This makes it a promising tool for biomedical applications such as cardiovascular flow systems.

\section{Acknowledgments}
This research was supported by the Platform for Advanced Scientific Computing (PASC, \url{http://
www.pasc-ch.org}) under the AV-FLOW project (\url{http://www.pasc-ch.org/projects/2013-2016/
av-flow}). The authors also acknowledge the support of the Swiss National Supercomputing Centre 
(CSCS, cscs.ch).

\emph{Declarations of interest: none}

\clearpage
\newpage

\appendix
\section{Analytical Inertia Benchmark}
\label{sec:inertiabench}
\begin{figure}[htb]
\begin{center}
\includegraphics[scale=0.30]{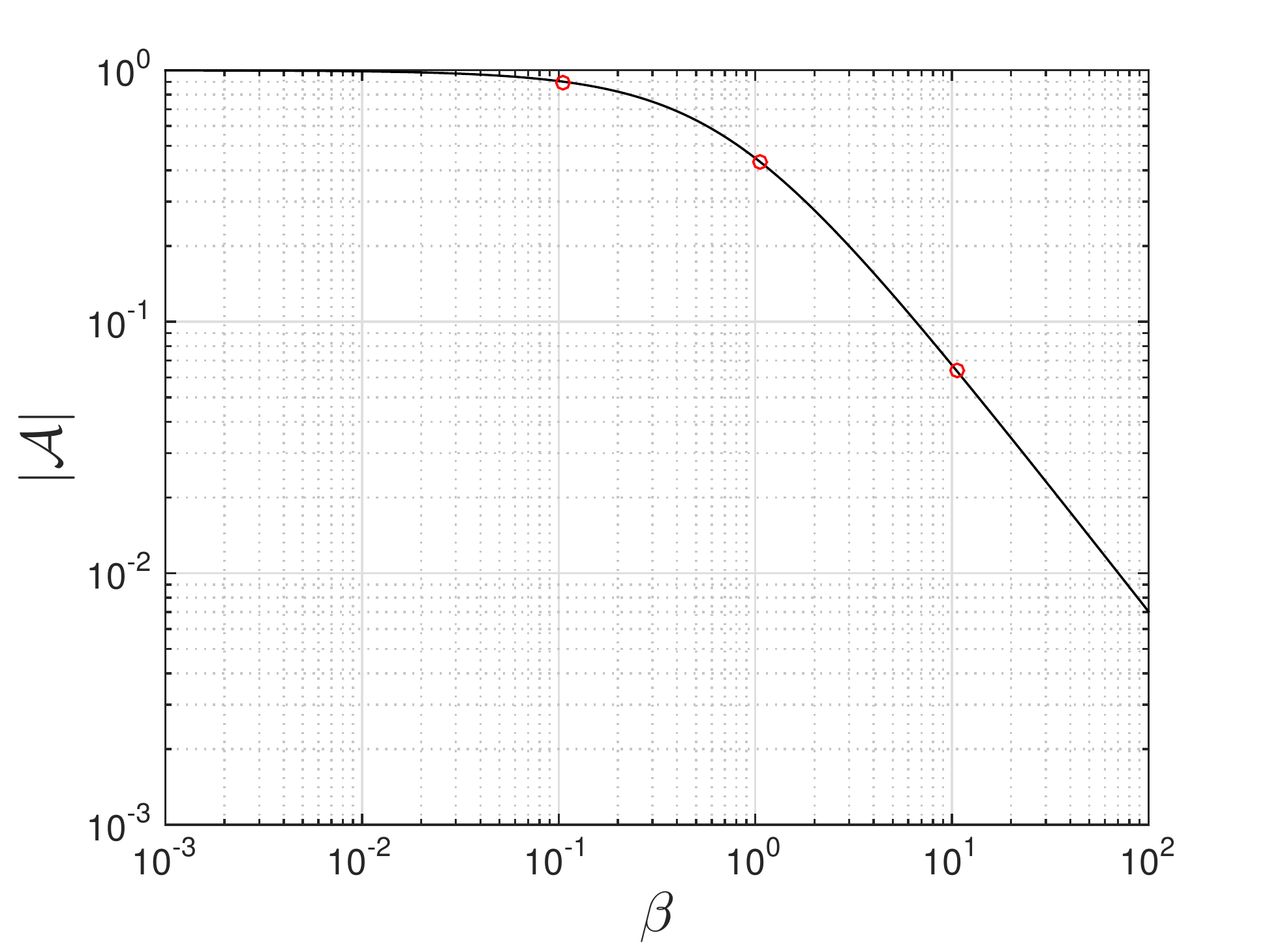}
\includegraphics[scale=0.30]{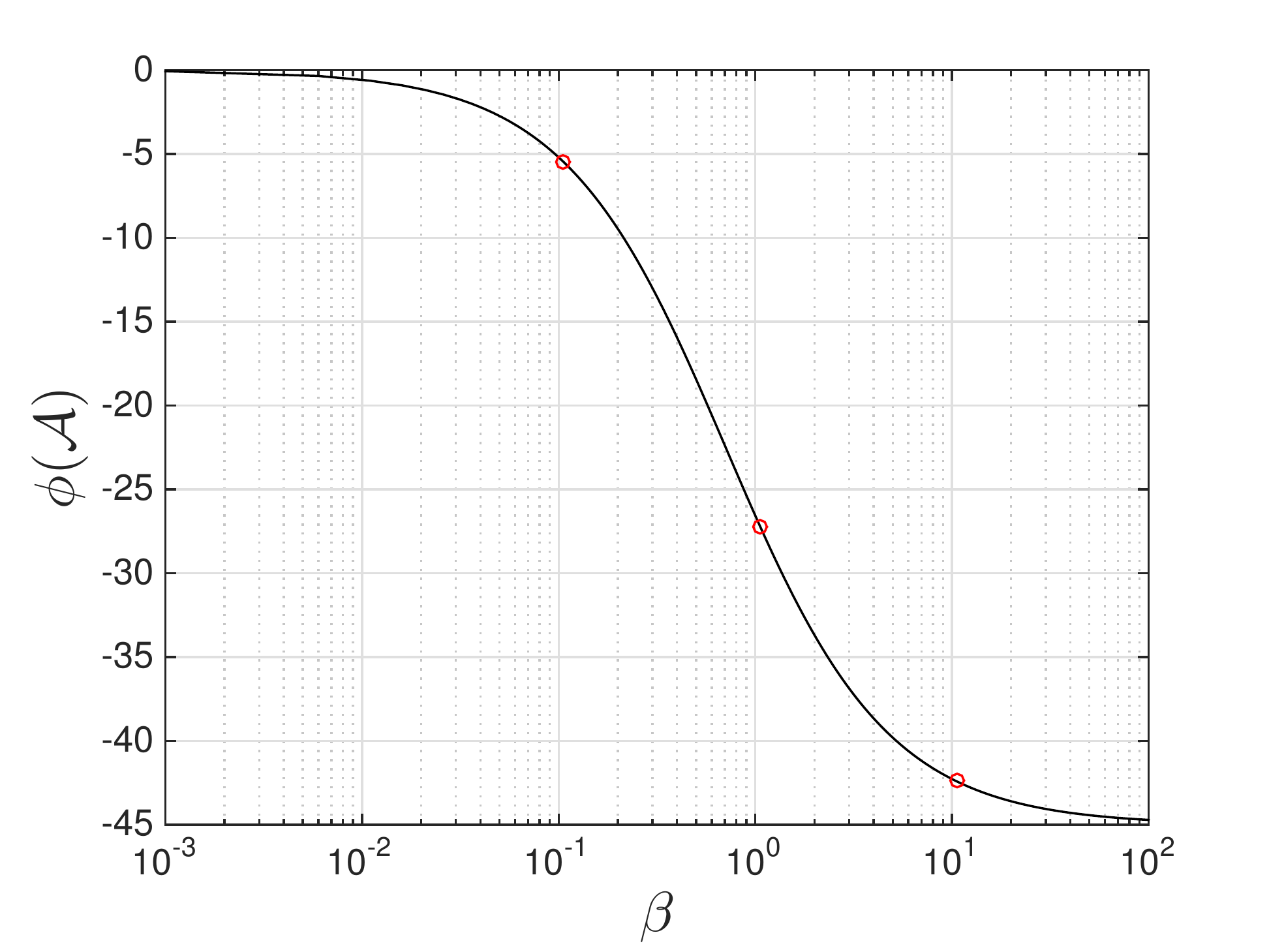}
\end{center}
\caption{Magnitude and phase diagram of the \textit{amplitude} function $\mathcal{A}(\beta)$. The red spots 
indicate the test cases analyzed in Section \ref{ssec:oscillation}. }
\label{fig09}
\end{figure}
We start considering the case of a  viscous fluid near a wall, driven by an oscillating pressure gradient.  With assumption of 
laminar flow and thus nearly parallel streamlines we can drop the advective term in the 
Navier--Stokes equations. 
Moreover, since the plate is supposed to be infinitely wide and long, the flow is invariant along the $x$-direction, i.e $
\frac{\partial (\cdot)}{\partial x}$ and from the continuity we see that  $\frac{\partial u}{\partial x}=0$. The 
unsteady Navier--Stokes equations  reduce to:
\begin{equation}
\label{ns_an}
\rho_f\frac{\partial v_f}{\partial t}=-\frac{\partial p}{\partial x} + \mu_f \frac{\partial^2 v_f}{\partial x^2}.
\end{equation}
The motion of the solid plate is forced by the viscous forces as follows:
\begin{equation}
2\delta_s \rho_s \frac{\partial v_s}{\partial t}=2\tau.
\end{equation}
Here $2 \delta_s$ is the thickness of the plate, and $\tau$ is the shear stress applied from the fluid to the solid.
Moreover, the following conditions have to be fulfilled on the fluid-structure interface:
\begin{align}
\label{bc_int}
\tau&=\mu\frac{\partial{v_f}}{\partial{y}},\\
v_s &=v_f.
\end{align}
By assuming the flow driven from a periodic pressure gradient, i.e. $\frac{\partial p}{\partial x}=-Ge^{j2 \pi f_0 t}$, 
we will have to choose an ansatz for $u(y, t)$ that is periodic as well and given the steady-state flow we can assume 
the frequency of the pressure profile as main frequency of the solution. Thus our ansatz becomes
\begin{align}
v_f(y,t) &=\mathcal{R}e\{V_f(y)e^{j 2\pi f_0t}\},\\
v_s(t)   &=\mathcal{R}e\{V_se^{j 2\pi f_0t}\}.
\end{align}
Plugging this into Equation~\eqref{ns_an}, assuming that the velocity is bounded for $y \rightarrow \infty$ and using the 
boundary conditions on the interface \eqref{bc_int} we get the following solution:
\begin{equation}
V_f(y)=\mathcal{R}e\left\{v_{f,\infty}\left(1-\frac{2\beta j}
{1+j+2\beta j}e^{(1+j)\alpha}e^{-(1+j)\alpha\frac{y}{\delta}}
\right)\right\},
\end{equation}
where  $j$ is the imaginary unit,  $v_{f,\infty}$ is the  \textit{free stream} 
velocity with 
\begin{align}
v_{f,\infty} & :=-\frac{jG}{2\pi f_0 \rho_f}, \\
\alpha         & :=\delta_s\sqrt{\frac{2 \pi f_0\rho_f}{2\mu}}, \\
\beta          &:=\frac{\rho_s}{\rho_f}\alpha \label{eq:beta}
\end{align}
and $G$ is the amplitude of the pressure gradient, $f_0$ is the frequency of the 
oscillating driving pressure gradient and $h=2\delta_s$ is the thickness of the solid plate.

The evaluation of the fluid velocity at $y=\delta_s$ gives the velocity of the solid plate which reads
\begin{equation}
V_s = V_f(y=\delta_s)=\mathcal{R}e\left\{v_{f,\infty}\left(1-\frac{2\beta j}{1+j+2\beta j}\right)\right\}.
\end{equation}
From here the  \textit{amplitude} ratio $\mathcal{A}:=v_s/v_{f,\infty}$ is derived, 
which gives (i) the amplitude of the velocity oscillation $v_s$ of the inert plate with respect to the amplitude of the 
oscillations of the  \textit{free stream} velocity  $v_{f,\infty}$  and (ii) the phase lag between the two, both depending on $\beta$ as 
follows :
\begin{equation}
\label{transfer_fun}
\mathcal{A}(\beta)=1-\frac{2\beta}{2\beta+1-j}.
\end{equation}
The magnitude $|\mathcal{A}|$ and the phase spectrum $\phi(\mathcal{A})$ of the \textit{amplitude} ratio  $\mathcal{A}(\beta)$ are shown in Figure~\ref{fig09}.

\section*{References}
\bibliographystyle{elsarticle-num}
\bibliography{JCP_methods}

\end{document}